\documentclass[preprint]{aastex}

\let\url\relax

\title{Measurement of $\Omega_{m}$, $\Omega_{\Lambda}$ from a blind analysis
of Type~Ia supernovae with CMAGIC:  Using color information to verify the 
acceleration of the Universe}
\shorttitle{Measurement of \om , \ol\ with CMAGIC}
\shortauthors{Conley \etal }
\author{ A.~Conley\altaffilmark{1,2,8},
G.~Goldhaber\altaffilmark{1,2},
L.~Wang\altaffilmark{1},
G.~Aldering\altaffilmark{1},
R.~Amanullah\altaffilmark{3},
E.~D.~Commins\altaffilmark{2},
V.~Fadeyev\altaffilmark{1},
G.~Folatelli\altaffilmark{4},
G.~Garavini\altaffilmark{5},
R.~Gibbons\altaffilmark{6},
A.~Goobar\altaffilmark{3},
D.~E.~Groom\altaffilmark{1},
I.~Hook\altaffilmark{7},
D.~A.~Howell\altaffilmark{8},
A.~G.~Kim\altaffilmark{1},
R.~A.~Knop\altaffilmark{6},
M.~Kowalski\altaffilmark{1},
N.~Kuznetsova\altaffilmark{1},
C.~Lidman\altaffilmark{9},
S.~Nobili\altaffilmark{5},
P.~E.~Nugent\altaffilmark{1},
R.~Pain\altaffilmark{5},
S.~Perlmutter\altaffilmark{1,2},
E.~Smith\altaffilmark{6},
A.~L.~Spadafora\altaffilmark{1},
V.~Stanishev\altaffilmark{3},
M.~Strovink\altaffilmark{1,2},
R.~C.~Thomas\altaffilmark{1}, and
W.~M.~Wood-Vasey\altaffilmark{1,2} \\
 (THE SUPERNOVA COSMOLOGY PROJECT) \\
}

\email{ conley@astro.utoronto.ca }

\altaffiltext{1}{E.\ O.\  Lawrence Berkeley National Laboratory, 
  1 Cyclotron Rd., Berkeley, CA 94720, USA }
\altaffiltext{2}{Department of Physics, University of California Berkeley, 
  Berkeley, 94720-7300 CA, USA}
\altaffiltext{3}{Department of Physics, Stockholm University,  
  Albanova University Center, S-106 91 Stockholm, Sweden}
\altaffiltext{4}{Observatories of the Carnegie Institution of  Washington,  
  813 Santa Barbara St., Pasadena,  CA 91101 }
\altaffiltext{5}{LPNHE, CNRS-IN2P3, University of Paris VI \& VII, Paris, 
  France }
\altaffiltext{6}{Department of Physics and Astronomy, Vanderbilt University, 
  Nashville, TN 37240, USA}
\altaffiltext{7}{Department of Physics, University of Oxford, Nuclear \& 
  Astrophysics Laboratory,  Keble Road, Oxford, OX1 3RH, UK}
\altaffiltext{8}{Department of Astronomy and Astrophysics, University of 
  Toronto, 50 St.\ George St., Toronto, Ontario M5S 3H4, Canada}
\altaffiltext{9}{European Southern Observatory, Alonso de Cordova 3107, 
  Vitacura, Casilla 19001, Santiago 19, Chile }

\newcommand{\dmofb}{\ensuremath{\Delta m_{15}\left( B \right)}}
\newcommand{\BmV}{\ensuremath{B-V}}
\newcommand{\Ebv}{\ensuremath{E(B\,-\,V)}}

\newcommand{\Bbvzs}{\ensuremath{B_{BV0.6}}}

\newcommand{\betabv}{\ensuremath{\beta_{BV}}}
\newcommand{\etal}{et~al.}
\newcommand{\om}{\ensuremath{\Omega_{m}}}
\newcommand{\ol}{\ensuremath{\Omega_{\Lambda}}}
\newcommand{\scriptm}{\ensuremath{\mathcal{M}}}
\newcommand{\scripte}{\ensuremath{\mathcal{E}}}
\newcommand{\al}{\ensuremath{\alpha}}
\newcommand{\chisq}{\ensuremath{\chi^{2}}}
\newcommand{\sg}{\ensuremath{\sigma}}
\newcommand{\sgi}{\ensuremath{\sigma_{int}}}
\newcommand{\mb}{\ensuremath{m_{B}}}
\newcommand{\rb}{\ensuremath{{\cal R}_B}}

\newcommand{\BmVBmax}{\ensuremath{(B-V)_{Bmax}}}
\newcommand{\EBmVBmax}{\ensuremath{E(B\,-\,V)_{Bmax}}}
\newcommand{\kcorr}{\ensuremath{K}-correction}
\newcommand{\kcorrs}{\ensuremath{K}-corrections}

\begin{document}

\begin{abstract}
We present measurements of \om\ and \ol\ from a blind analysis of 21
high-redshift supernovae using a new technique (CMAGIC) for fitting
the multi-color light-curves of Type~Ia supernovae, first introduced 
by \citet{Wang:03}.  CMAGIC takes advantage of the remarkably simple 
behavior of Type Ia supernovae on color-magnitude diagrams, 
and has several advantages over current techniques based on maximum
magnitudes.  Among these are a reduced sensitivity to host galaxy dust
extinction, a shallower luminosity-width relation, and the relative
simplicity of the fitting procedure. This allows us to provide a cross-check 
of previous supernova cosmology results, despite the fact that 
current data sets were not observed in a manner optimized for CMAGIC.
We describe the details of our novel blindness procedure, which is designed
to prevent experimenter bias.  The data are broadly consistent with
the picture of an accelerating Universe, and agree with a flat
Universe within 1.7 \sg , including systematics.
We also compare the CMAGIC results directly with those of maximum 
magnitude fits to the same supernovae, finding that CMAGIC favors more 
acceleration at the 1.6 \sg\ level, including systematics and the 
correlation between the two measurements.  A fit
for $w$ assuming a flat Universe yields a value that is
consistent with a cosmological constant within 1.2 \sg .
\end{abstract}

\keywords{cosmological parameters --- cosmology: observations ---
 supernovae: general}

\section{INTRODUCTION}
\nobreak
Type Ia supernovae (SNe~Ia) have proved to be an extremely valuable
tool for measuring the cosmological parameters, as they are the best 
high-luminosity standard candles currently known to astronomy.
Studies of the peak $B$-band luminosities of high redshift SNe~Ia
led to the surprising discovery by two independent groups (the Supernova
Cosmology Project (SCP; Perlmutter \etal\ 1998, Perlmutter \etal\ 1999
(hereafter P99))  and the High-z Supernova Search Team (HZSST;
Garnavich \etal\ 1998; Schmidt \etal\ 1998; Riess \etal\ 1998),
that the expansion of the Universe is accelerating.  This acceleration is
consistent with some form of `dark energy', possibly Einstein's cosmological
constant $\Lambda$.  The implications of this result for the future fate of 
the Universe and our understanding of fundamental physics are profound;
therefore, it is extremely important that it be verified by
independent methods.

The best approach is to make use of alternative measurements that depend
on other physical processes.  There are now several additional lines
of evidence that support the accelerating Universe, but most are based on 
combining several different measurements.  For example, the combination 
of the angular size of 
fluctuations on the surface of last scattering of the cosmic microwave
background (CMB) with measurements of the clustering of mass on large
scales \citep{Spergel:03,Tegmark:04,Eisenstein:05} provides strong 
evidence for a  dark energy component.  There is also a direct detection 
of dark energy using the integrated Sachs-Wolfe effect \citep{Padmanabhan:05}.
It is encouraging that these different lines of evidence, 
which depend on very disparate physical processes and probe very 
different cosmic epochs, are consistent with a $\om \sim 0.3$, $\ol \sim 0.7$ 
Universe.

Still, SNe~Ia provide the best direct evidence for dark energy, and any
improvement in our understanding of their properties is very welcome.
There are several possible alternative explanations for the SN result.
Since dark energy 
manifests itself in this context as high-redshift SNe~Ia being slightly dimmer 
than expected, the most obvious alternative explanation is that this dimming 
is caused by extragalactic dust,  either in intergalactic space or in the host
galaxies of the SNe.  Another possibility, and a
significantly more difficult one to quantify, is that high redshift SNe
are somehow dissimilar from low redshift SNe in a way that we have not
yet detected.  This paper presents results based on an analysis of SNe~Ia
with a new method (CMAGIC, for Color-MAGnitude Intercept Calibration) 
introduced in \citet{Wang:03} (hereafter W03) that partially addresses both 
issues.

There is no unique choice for the magnitude to associate with an SN~Ia
because their luminosity varies in time.
For convenience, virtually all previous studies have used the $B$ magnitude 
at maximum brightness, \mb , as the standardized candle,
but there is no {\it a priori} reason why this choice is optimal.
\mb\ is generally determined by fitting an empirical curve to the $B$-band
brightness as a function of time and reading off the peak value.
When available, observations in other passbands are frequently 
incorporated into the fitting procedure.  There is a well-established
empirical relation between absolute magnitude and the width of the light curve 
as parameterized by stretch (Perlmutter \etal\ 1997; P99; 
Goldhaber \etal\ 2001), \dmofb\ \citep{Phillips:93,Phillips:99} or 
the MLCS parameter $\Delta$ \citep{Riess:96} in the sense that SNe with wider, 
more slowly declining light curves (high stretches) are intrinsically brighter. 
Here the stretch parameterization is used.

Since ordinary interstellar dust both extinguishes and reddens light, P99
compared the distributions of \BmV\ colors at maximum luminosity of the 
low and high redshift SN samples, finding no significant evidence that the 
high redshift sample is more reddened. It should be emphasized that the 
SN measurement of \om\ and \ol\ is relative -- as long as the low-
and high-redshift samples suffer the same amount of extinction (or any other
bias), there is no effect on the final result. \citet{Sullivan:02}
decomposed the SN sample into subsets based on the Hubble 
type of their host galaxies, a powerful approach because early-type galaxies
are expected to have little or no dust, and found that \ol\ was detected
in each subsample.  A difficulty with 
this analysis is that the resulting error bars on \om, \ol\ are necessarily 
much larger because the morphological subsets have considerably 
fewer SNe than the full sample.

One may attempt to measure the reddening for each SN by measuring 
its color and correcting for host galaxy extinction by assuming a dust 
extinction law.  The error in the extinction correction usually dominates
the statistical errors of each SN.  In early work the HZSST
team made use of an asymmetric prior on the intrinsic extinction 
distribution to limit the propagated uncertainties resulting from the 
extinction correction \citep{Riess:98} while performing light-curve
fits, which can bias the results under some circumstances (P99).  
More recent papers have made improvements in the form of the prior and
its application and corrected this problem \citep{Barris:04, Riess:04},
although at the potential cost of enhanced sensitivity to any evolution in
the extinction distribution.  \citet{Knop:03} (hereafter K03) made use 
of high quality color measurements made possible by the 
{\it Hubble Space Telescope (HST)} to estimate the extinction values 
of individual SN without making use of such a prior.

The evolution issue is extremely difficult to address.   To first
order evolution should not be a concern because the diversity of the
environments in which local SNe~Ia occur is much larger than the mean
difference in environment between the high and low redshift samples.  
While there are some properties of SNe~Ia that are known to correlate with host
environment, these correlations disappear once the width-luminosity
relation is taken into account \citep{Hamuy:00}.  The analysis of
\citet{Sullivan:02} also has relevance for this question because it compares
SNe~Ia from similar host environments at high and low redshift.  One can 
also compare individual SNe in more detail spectroscopically 
\citep{Hook:05}, although such measurements are taxing even for modern
\mbox{8-10m} class telescopes.  In a spectroscopic study of 12 high 
redshift SNe, \citet{Garavini:05} found no evidence for evolution.

CMAGIC offers some benefits with respect to dust and evolutionary models,
as described further in \S \ref{subsec:dust} and \S \ref{subsec:evolution}.
It is possible to define a standard candle magnitude with CMAGIC, 
and because of the nature of the CMAGIC relationships, this magnitude is
affected by the same amount of dust by roughly half as much as \mb . On 
the evolutionary front the situation is more complicated.  There are some
potential evolutionary effects for which CMAGIC offers advantages, but
it is uncertain how important this is because the effects of these theories 
have not been clearly delineated. Because CMAGIC depends on light-curve
data in a very different fashion than maximum magnitude fits, and in
particular because it is much more sensitive to later epochs relative
to maximum light, for some potential evolutionary effects we can expect
the CMAGIC magnitude to be affected differently. However, this is difficult
to quantify given the current lack of detailed predictions from theories
of SN evolution.  Combining these two considerations, 
CMAGIC can provide a powerful cross check of previous SNe~Ia cosmology
results.  Because we are attempting to verify previous results, it is
important to prevent the analysis from being unintentionally biased towards
the expected outcome.  To this end a blindness technique has been
developed and used during the cosmological analysis in this paper.

Perhaps for some of the above reasons, low redshift SNe~Ia analyzed 
with CMAGIC have a smaller intrinsic variation than the maximum 
magnitudes of the same SNe without extinction correction ($\sgi= 0.12$ 
mag, compared with approximately $0.17$ mag for \mb ).  For many current
data sets, the intrinsic variation dominates over observational errors, so
CMAGIC may allow us to obtain tighter constraints on the cosmological
parameters for a similar observational expense in future surveys.

The goals of this paper are twofold:  (1) to show that the CMAGIC
relations hold at high redshift for well measured SNe, and (2) to
measure the cosmological parameters from already existing data sets and
use this to cross-check previous results.   We first describe CMAGIC in 
more detail (\S \ref{sec:cmagic}).  We then describe the data sample
(\S \ref{sec:data}) and the CMAGIC fitting procedures (\S \ref{sec:cmagfits}), 
and then we use these to demonstrate that CMAGIC works for high redshift 
SNe  (\S \ref{sec:highzdemo}).  Once this is established, we proceed to the 
primary analysis of this paper, the cosmological fits.  First we describe 
the cosmological fitting techniques (\S \ref{sec:cosfits}), including a 
discussion of the blindness technique (\S \ref{subsec:blindness}).
Finally, the cosmological results are presented (\S \ref{sec:cosresults}),
systematic effects are discussed (\S \ref{sec:systematics}), and
the results are analyzed (\S \ref{sec:analysis}).

\section{CMAGIC}
\label{sec:cmagic}
\nobreak
CMAGIC is described in considerably more depth in W03.  Here we
provide a brief review of the relations, define the magnitude (\Bbvzs ) 
used in this study, and discuss the benefits of CMAGIC with respect to 
extinction and evolution.

\subsection{CMAGIC Relations}
\label{subsec:cmagrelations}
\begin{figure}
\plotone{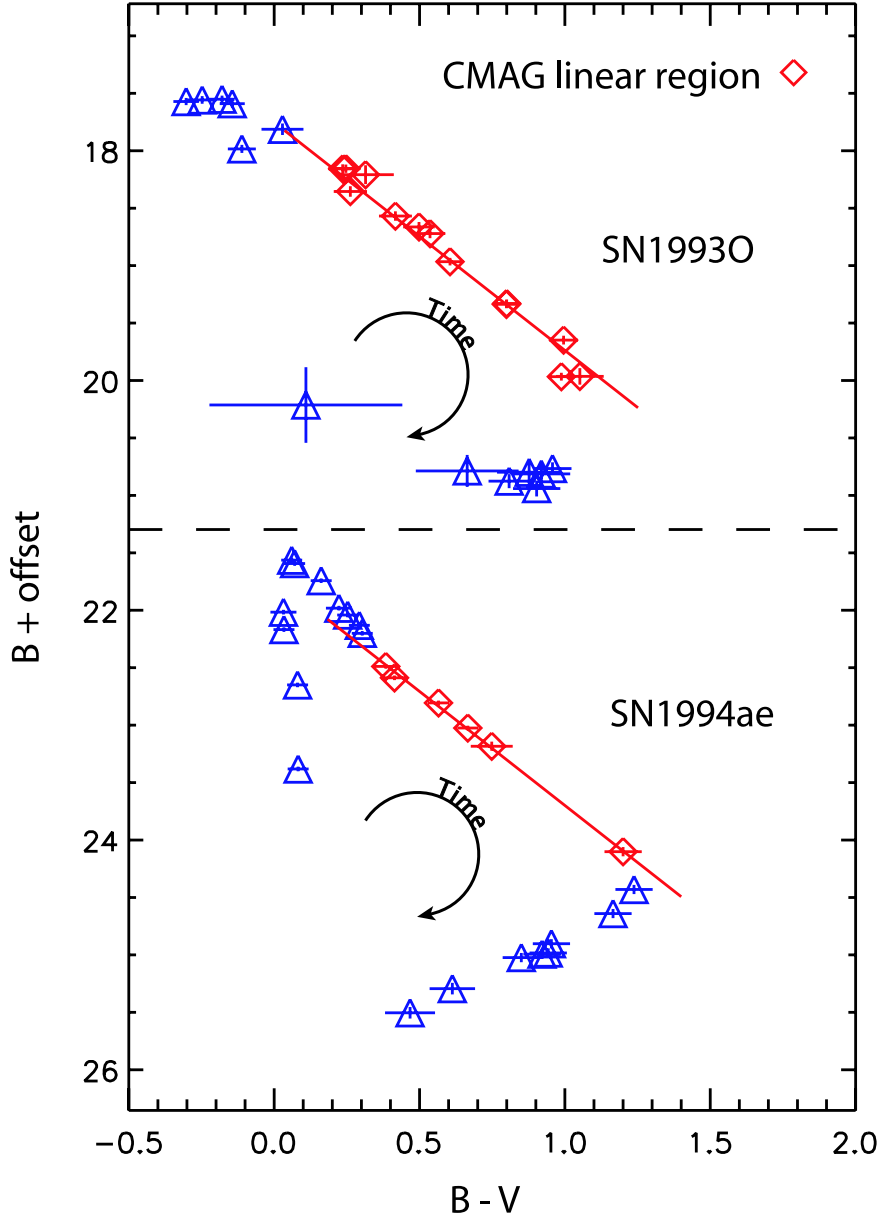}
\caption{$B$ vs. \BmV\ CMAGIC diagrams for SN1993O, a fairly typical low 
 redshift SN Ia (z = 0.052, stretch = 0.927), and SN1994ae (z = 0.004, 
 stretch = 1.006).
 The points in the linear region (based on the date relative to maximum light) 
 are shown as diamonds for both SNe. The slopes were fixed at \betabv\ = 1.98 
 for this fit. SN1994ae displays a bump feature prior to the linear region.
 \label{fig:cmag}}
\end{figure}

\nobreak
CMAGIC is based on the behavior of SNe~Ia in color-magnitude diagrams.
Starting approximately 1 week after $B$ maximum and lasting approximately
3 weeks, the relation between the $B$ magnitude and \BmV\ color is
strikingly linear.   This holds true for other colors as well (at least $B-R$, 
$B-I$). Some typical low redshift examples are shown in figure~\ref{fig:cmag}. 
The temporal extent of this linear region is a function of stretch, with 
slower, higher stretch light-curves starting and ending their linear behavior 
later. The slope, $\beta$, of the linear region has a narrow distribution.  
Currently very few rest-frame $R$ and $I$ observations are available for high 
redshift SNe~Ia, so here we consider only $B$ versus \BmV .  The simplicity of 
this behavior is so far not completely explained by theory, which gives it a 
status similar to the empirical width-luminosity relation.  Prior to the linear
region, the majority of SNe~Ia are less luminous than the linear extrapolation.
However, a minority (typically those with high stretch) display excess 
luminosity, which is referred to as a `bump'.  Standard light-curve template 
fitting techniques (stretch, MLCS) do not adequately reproduce the CMAGIC 
relations. Both issues are discussed in more detail in W03.

\begin{figure}
\plotone{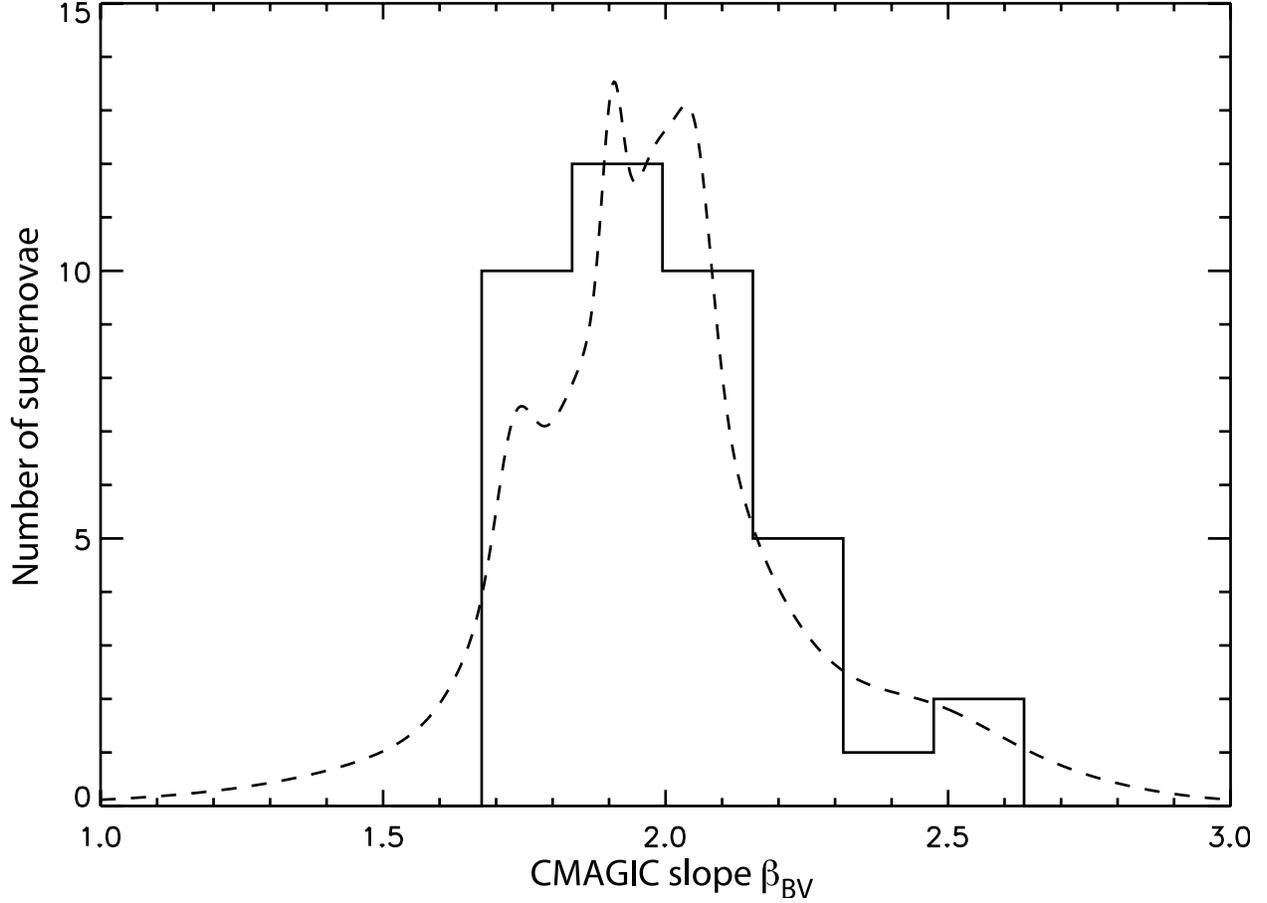}
\caption{Histogram of \betabv\ for the 44 low redshift SNe~Ia used in the
 primary fit,  after \kcorr\ (solid line).  Overlain is an ideogram
 of the same distribution (dashed line). The ideogram is constructed by 
 adding a Gaussian of the appropriate width and mean value for each SN,
 representing the best fit value of \betabv\ and its associated measurement
 error of each object.  This shows how the size of the measurement errors
 are affecting the distribution.  The mean of this distribution is 1.99 and 
 the RMS is 0.16. The binning is purely for display purposes. 
 \label{fig:slopes}}
\end{figure}

The distribution of slopes in this linear region is fairly narrow, with 
$\left< \betabv \right> = 1.98$ and a RMS of 0.16, as shown in 
figure~\ref{fig:slopes} for low-redshift SNe~Ia.  To first order, \betabv\ 
is affected by \kcorrs\ but not by extinction.  W03 explored fixing the 
slope at the mean value for all fits.  The effects of this assumption are 
quite minor, but it is possible to improve on this procedure by including 
information about the distribution of slopes in the fitting procedure 
(\S\ref{sec:cmagfits}).

The CMAGIC relation for $B$ versus \BmV\ can be written conveniently in the form
\begin{equation}
B = \Bbvzs + \betabv \left( B - V - 0.6 \right),
\end{equation}
which defines \Bbvzs\ as the $B$ magnitude when $\BmV = 0.6$; this is the 
magnitude used as a standard candle in this paper.  The particular \BmV\ 
color is chosen to minimize the covariance between the standard candle 
magnitude and the slope \betabv , as it is approximately the mean \BmV\ 
color in the linear region of an unextinguished SN~Ia. Because the color 
roughly measures the ejecta temperature, by evaluating the
magnitude at a fixed color we essentially ensure that all SNe are evaluated
at a point where their physical properties are similar.  

The behavior of an SN Ia on a CMAGIC diagram can also be viewed temporally.
Proceeding in a clockwise fashion around the curves in figure~\ref{fig:cmag},
a typical, unextinguished SN Ia usually has a color of approximately $\BmV = 0$
at maximum, and evolves rapidly to the red for about a month.  After this it 
enters the so-called nebular phase and evolves bluewards, again in a linear 
fashion.  This second linear region has some interesting properties, but 
since data at such late epochs are very rarely available for high-redshift 
SNe, we do not discuss it further here.  With good time coverage it is possible
to determine the extent of the linear region by examination, but this is
generally not possible with current high redshift data.  Fortunately, the
beginning and ending dates of the linear region relative to the date of
$B$ maximum  form a well-defined sequence in terms of stretch and the presence
or absence of the bump feature.  Using well-observed low-redshift SNe to
determine the earliest and latest points in the linear region as a function
of stretch, we find that the beginning date of the linear region is well 
described by $t_b = 5 + 3 \left(s - 1\right)$ and the ending date by 
$t_e = 29 + 40 \left(s - 1\right)$, where both are measured in 
rest-frame days relative to $B$ maximum and $s$ is the stretch.  SNe~Ia with 
bumps (e.g., the lower panel of figure~\ref{fig:cmag}) do not fit smoothly into
this scheme and are well represented by $t_b = 13.5$ and $t_e = 30$.  This
suggests a possible source of bias in the analysis of the high redshift sample,
since the presence or absence of a bump may be difficult to detect given 
the typical quality of high redshift photometry.  Fortunately, for this data 
sample this issue proves to be unimportant (Appendix~\ref{apndx:bumps}).

Detailed studies (Appendix~\ref{apndx:correlations}) show that the fitting 
procedure induces weak negative correlations between \Bbvzs\ and \mb , 
at least for current light-curve templates.  Clearly, these templates have 
missed some aspect of SNe~Ia behavior (or the correlations would be much 
stronger), and \Bbvzs\ provides some additional information that can be 
used to constrain the cosmological parameters.  Peculiar velocities, stretch 
correction, and extinction induce additional correlations between these 
magnitudes.

\subsection{Host Galaxy Dust}
\label{subsec:dust}
\nobreak

Interstellar dust is a major component of our and other galaxies.  
A good review can be found in \citet{Draine:03}. Ordinary dust both 
extinguishes and reddens starlight because it absorbs blue light more 
strongly than red light.  The relative amount of absorption between wavelengths
is characterized by an absorption law such as that of \citet{Cardelli:89}.
For an object with a stellar spectrum, the extinction in the $B$-band $A_B$ 
(in magnitudes) is related to the amount of reddening \Ebv\ by 
$A_B = \rb \Ebv$.  For SNe, which do not have stellar-like spectra, 
and whose spectral features change with time, this is not strictly appropriate,
but \rb\ is still useful as a parameterization of the extinction law.  A 
typical value in our Galaxy is $\rb = 4.1$, although it varies considerably 
along different lines of sight \citep{Fitzpatrick:99}.  The characteristic 
scatter of \rb\ is not well constrained.  

So far it has not been feasible to measure the extinction law directly
for the host galaxies of high redshift SNe, so the general approach has been
to assume that the \rb\ values for the high and low redshift SNe samples
are identical.  This assumption takes several forms.  In the primary fit
of P99 (fit C) no extinction correction is performed, but it is argued that 
the similarity of the observed \Ebv\ distributions of the two samples
implies that host  galaxy dust extinction is not contaminating the 
cosmological results.  Because \rb\ is necessary to transform \Ebv\ into 
the amount of extinction, this is tantamount to assuming that \rb\ is the 
same for the two samples.  There is a theoretical and empirical 
expectation that the SN sample suffers from relatively little 
extinction \citep{Hatano:98}.  K03 perform an extinction correction by 
comparing the measured \BmV\ at maximum to an empirical model, then 
converting this to $A_{B}$ by assuming a value for \rb .
\citet{Riess:98,Riess:04,Tonry:03,Barris:04} use a similar procedure.
Previous analyses have generally performed a color cut on their SN
samples on the theory that large color excesses may represent SNe in dustier 
environments where the value of \rb\ is likely to depart from the fiducial 
value.  It is interesting to note that we may now have evidence
for higher mean extinction at high redshift.  The recent SN sample
of \citet{Riess:04}, which represents the deepest, highest redshift SN survey 
yet published, has much higher host galaxy extinction values than any other 
available SN sample, although survey selection effects may explain this
result.  

\begin{figure}
\plotone{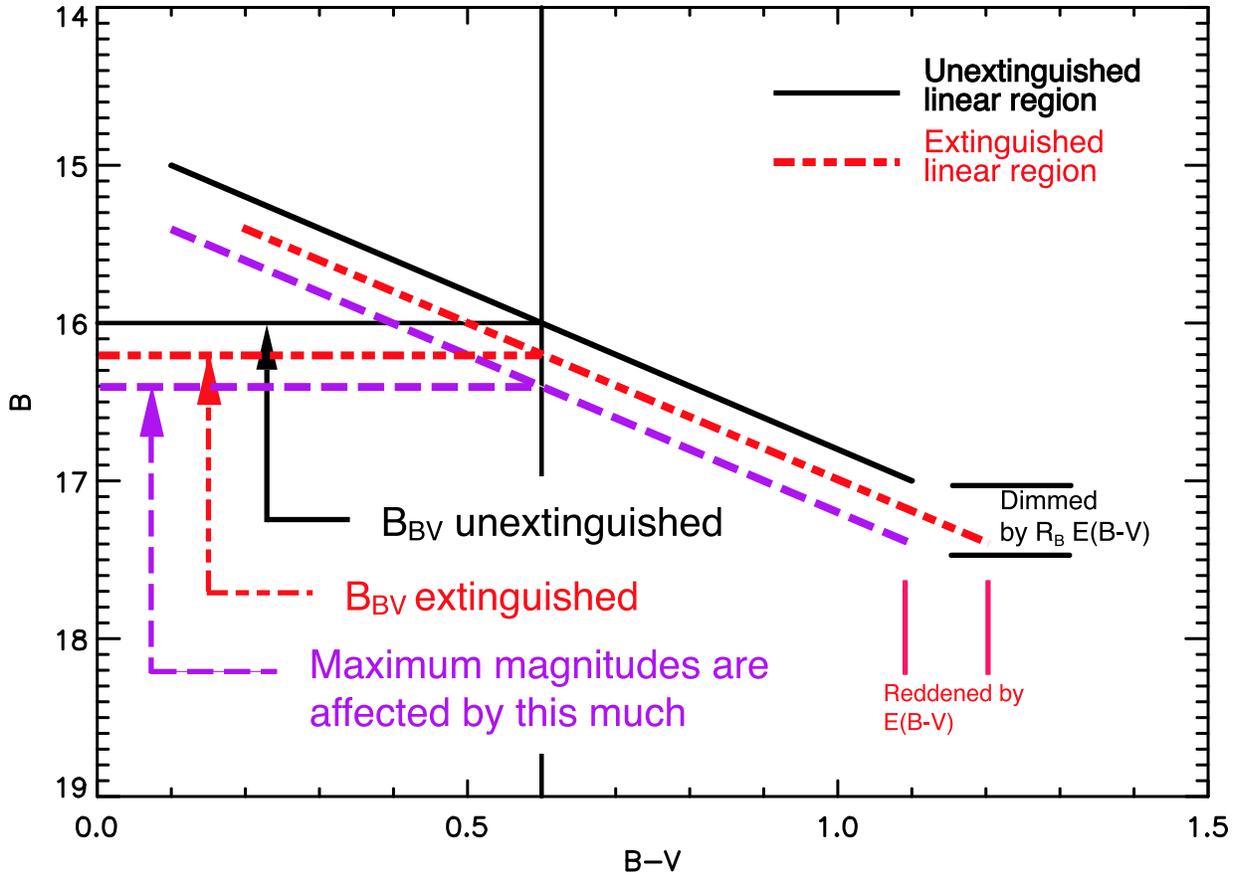}
\caption{Schematic representation of how host galaxy dust affects \Bbvzs .
 Ordinary dust both extinguishes and reddens light.  Here
 an extinction of $A_B = 0.4$ mag is shown.  The solid line represents
 the unextinguished linear region.  The bottom line represents the effects
 of extinction without reddening, the middle line includes both the dimming
 and reddening effects.  The critical point is that, because the linear
 relation is always evaluated at the same color to form \Bbvzs , 
 the two effects partially cancel. \label{fig:cmagdust} }
\end{figure}

Because of the nature of the linear CMAGIC relations, the effective 
${\cal R}$-value for \Bbvzs\ is approximately half of the value that it takes 
for \mb\ (assuming a standard dust law), as shown schematically in 
figure~\ref{fig:cmagdust}.  The critical point is that the magnitude
is always evaluated at the same fixed color, and therefore the extinction
and reddening effects partially cancel.  Since SNe~Ia redden as they 
evolve along the linear relation,  ${\cal R}_{\Bbvzs} = \rb - \betabv$.  
For normal dust, \Bbvzs\ is less affected than \mb , which results in 
smaller uncertainties arising from the extinction correction, if a fixed 
\rb\ is assumed.  Because the boundaries of the linear region are determined
by date relative to maximum and not color, \Bbvzs\ remains less affected
even if the amount of extinction is large enough that $\BmV = 0.6$ does
not lie within the linear region.  The precise epoch of maximum light is 
not nearly as important for \Bbvzs\ as it is for \mb\ because the `roll-off' 
at the edges of the linear region is much less severe than it is near peak 
luminosity. Note that CMAGIC offers no benefits with respect to an evolving 
\rb\ -- the derivatives of \mb\ and \Bbvzs\ with respect to \rb\ are 
identical.  Nor does it offer any advantages for the so-called 'gray dust' 
($\rb = \infty$) suggested by \citet{Aguirre:99}.  Constraints on gray dust 
have been explored by \citet{Riess:00, Riess:04}, but also see 
\citet{Nobili:03, Nobili:05}.

Since \Bbvzs\ and \mb\ are affected by extinction differently, it is
possible to estimate the amount of extinction by comparing the
two magnitudes using the quantity \scripte , which is an estimator
of \mbox{ \Ebv } :
\begin{equation}
 \scripte = \frac{ \mb - \Bbvzs }{ \betabv } + \mbox{const}.
\end{equation}
Using this correction substantially increases the correlations 
between \mb\ and \Bbvzs . Assuming a standard extinction law (\rb = 4.1), 
the correlation coefficient between these two magnitudes climbs to 
$\rho > 0.7$ from $\left< \rho \right> = 0.15$ 
(Appendix~\ref{apndx:correlations}), even in the absence of significant 
extinction.  For this reason, this approach is not followed here.  However, 
for smaller values of \rb , such as those found by \citet{Tripp:99} and 
\citet{Guy:05}, this correlation is significantly reduced.

\subsection{Evolution of SNe~Ia}
\label{subsec:evolution}
\nobreak
The possibility that the average properties of SNe~Ia have evolved
between the current epoch and a redshift of 1 is of considerable 
concern for SN cosmologists.  So far it has been impossible 
to demonstrate conclusively that evolution is not the cause of the claimed 
cosmological results. The best that can be done is to continue to 
quantitatively add ``to the list of ways in which they are similar 
while failing to discern any way in which they are different'' 
\citep{Riess:99b}.  One method to approach this problem is to compare 
high and low redshift SNe in similar environments, as in \citet{Sullivan:02}, 
where we found no evidence for evolutionary biases. Since all
measured dependencies of SN Ia properties on local environment disappear
after stretch correction, and because of the diversity of environments in which
local SNe~Ia occur, concerns about evolution can be usefully restricted
to mechanisms that affect the width-luminosity relationship.

There are several theoretical models that predict possible avenues for
evolution.  \citet{Dominguez:01} and \citet{Hoflich:00} have investigated the 
effects of decreasing metallicity and changing progenitor mass on SN~Ia 
properties by constructing models of the progenitor star and then following 
them through detonation.  If $\Delta$ is the change in \BmV\ they
find that decreasing metallicity causes an SN to become slightly bluer 
($\Delta = -0.05$ for an extreme case) without affecting the maximum 
$B$ magnitude.  Most extinction corrections compare observed
colors to empirically derived color relations to calculate the 
amount of extinction.  If the intrinsic colors change, then the extinction 
correction will be incorrect.  If no extinction correction is applied,
then \mb\ is unaffected, while \Bbvzs\ is overestimated by
$\betabv \Delta \sim 2 \Delta$.  If an extinction correction is applied, 
then for positive values of $\Delta$, the extinction correction for \mb\ is
overestimated and the SN is assigned an extinction-corrected magnitude that
is too bright by ${\cal R}_{B} \Delta \sim 4 \Delta$.  \scripte , by contrast,
is underestimated, so once this correction is applied, \Bbvzs\ is too
dim by $\betabv \Delta - \left( \rb - \beta \right)^2 \Delta / \betabv$.
For typical values of \betabv\ and \rb , this cancels, and the extinction
corrected value of \Bbvzs\ is unaffected by this evolutionary effect.
In other words, either with or without extinction correction this particular
evolutionary model will have different effects on \mb\ and \Bbvzs , so
by comparing the two magnitudes this model can be evaluated against data.
We note that the range of metallicities considered in this study is
far greater than the expected change out to $z \sim 1$.

\section{DATA}
\label{sec:data}
\nobreak
Currently available SN data sets have not been observed in a 
manner optimized for CMAGIC, particularly at high redshift.  Out
of the roughly 100 SNe at $z > 0.1$ with light curves available
in the literature, only approximately 20 are useful for
CMAGIC purposes.  High redshift SNe are frequently not observed 
in the rest-frame $V$.  Even when such observations do exist, they are 
usually only intended to establish the color at maximum for the purposes of 
applying an extinction correction, and therefore are usually concentrated too 
close to the peak to lie within the CMAGIC linear region.  Future high redshift
data sets (SNLS \citep{Astier:05}, ESSENCE \citep{Matheson:05}, 
SDSS Supernova Search \citep{Sako:05}, SNAP \citep{Aldering:04}, 
LSST \citep{Pinto:04}) will not suffer from this limitation, as they are 
designed to obtain multi-color photometry for almost all observed epochs.
The current situation is considerably better 
for low redshift data sets, as many of these SNe have excellent multi-color
coverage.  There is an observational
cost associated with CMAGIC because the linear region is $\sim 1.2$ mag
dimmer than at peak, so the photometric error bars are larger for the same
observational effort.  Whether or not this extra cost is outweighed by
the benefits with respect to dust and/or evolution depends on the specifics
of the survey design.

We have attempted to construct a data sample including all SNe~Ia with 
published light curves.  
In order to eliminate SNe that cannot be useful for the purposes of 
this paper, we enforce the following requirements.  First, an object
must be at least plausibly an SN~Ia based on either light-curve shape,
spectroscopic ID, or host galaxy morphology.  Second, it must have at 
least one rest-frame \BmV\ observation.  For this purpose we require that 
the central wavelength of the redshifted $B$- or $V$-band lie within one 
HWHM of the central wavelength of the observed filter, which improves the
reliability of the \kcorrs\ by limiting the amount of extrapolation.
We also do not include 
observations taken with extremely wide filters, such as F110W and 
F160W NICMOS filters on {\it HST}.  These filters are wide enough that 
for many of the redshift ranges of interest they overlap considerably with
both $B$ and $V$ (and sometimes $R$), making it difficult to measure 
\BmV\ in a fashion that is not heavily influenced by the model used to 
calculate the \kcorrs .  Clearly it must be possible to
use these data in some fashion for CMAGIC, but it will require extreme care.
Observations in $B$ and $V$ are only 
combined to form \BmV\ if they are within 0.5 rest frame days of each other; 
the analysis is quite insensitive to this value.

This results in a sample of 131 SNe, of which one third are at redshifts greater 
than 0.3.  Note that we have not yet required that the \BmV\ point lie in the 
CMAGIC linear region, since this depends on the measured value of the stretch 
and date of maximum, or that the SN lie in the Hubble flow.  
The high-redshift portion of the sample comes from a fairly 
diverse set of sources.  There are 14 from P99,  six from K03, 
two from \citet{Garnavich:98}, one from \citet{Schmidt:98}, five from 
\citet{Riess:98}, four from \citet{Tonry:03}, 13 from \citet{Barris:04}, 
and one from \citet{Riess:04}.  The low-redshift sample is even more diverse,
but primarily comes from three sources: \citet{Hamuy:96}, 
\citet{Riess:99a} and \citet{Jha:05}. Source information is provided in 
tables~\ref{tbl:primarysamplowz} and \ref{tbl:primarysamphighz}.
Once a reasonable series of cuts are applied to this sample 
(\S \ref{subsec:cuts}), approximately half of the SNe remain and are 
used in the cosmological analysis.  

\section{CMAGIC FITTING PROCEDURES}
\label{sec:cmagfits}
\nobreak
In order to determine if an individual data point lies within the linear
CMAGIC region for a particular SN it is necessary to know 
the stretch and the date of $B$ maximum, although not to a high degree
of accuracy.  These are determined by performing a template fit to the $B$ 
and $V$ light curves in a manner similar to P99 and K03.  Briefly,
light-curve fits are performed using a \chisq\ minimization procedure based
on MINUIT \citep{James:75} with both \kcorrs\ and corrections for
Milky Way dust extinction taken into account.  The light-curve template
is that of K03 (which uses the $B$ template of \citet{Goldhaber:01} 
but a different $V$ template).  For the photometry from P99 and K03,
the photometric correlation matrices were used in the light-curve fits.  
These reflect
the correlations between different observations of the same SN
induced by the subtraction of the final reference image(s).
For the literature objects, where this information was not available,
the observations are assumed to be uncorrelated.
In order to prevent systematic errors arising from 
differences in fitting procedures, we have only 
included SNe that we can treat consistently, i.e.\ with our own light-curve 
fitting procedure and \kcorrs .

The correlation of the bump feature with different $B$ and $V$ stretch values
complicates matters.  As explained in Appendix~\ref{apndx:bumps},
SNe~Ia with bumps can be fitted by the standard stretch templates if
the ratio between $B$ and $V$ stretch values is allowed to vary.
In order to handle this situation, three light-curve fits
were performed for each SN -- joint $B$ and $V$, $B$ only, and $V$ only. 
In joint fits the dates of maximum and stretch values of the two filters
are fixed relative to each other by the light-curve template.  Except
when a bump is visible in the CMAGIC diagram, the joint fit is used.  
The reduced detectability of the bump feature at high redshift
due to reduced data quality is a concern that is further discussed in 
Appendix~\ref{apndx:bumps}.

\kcorrs\ play a critical role in this procedure.  At high redshift
cross-filter corrections are necessary \citep{Kim:96}, but even at 
low redshift same-filter \kcorrs\ are not insignificant.  Erroneous 
\kcorrs\ alter the slope of the CMAGIC linear region, 
unlike extinction.  Those used in this paper are based on the prescription of
\citet{Nugent:02} but with the time series of spectral templates and 
empirical stretch-color relation of K03. 
Milky Way extinction is included in this calculation using 
the dust map of \citet{Schlegel:98}.  Our approach naturally takes into
account the non-stellar nature of SN spectra and their variation
with epoch.  Errors associated with the \kcorrs\ are discussed in
\S\ref{sec:systematics}, where we also discuss the effects of several 
other modifications to the fitting procedure described here.

Since the \kcorr\ is a function of stretch and epoch, the light-curve
fits must be performed in an iterative manner.  On the first iteration the
stretch is set to 1 and the date of maximum is set to the date of the
brightest point.  The combined Milky Way and \kcorrs\ are calculated
and the light curve is fitted, and the new stretch and date of maximum
are used to calculate new corrections.  This process is iterated until 
convergence.  The majority of SNe converge
within three iterations, but the maximum number allowed is 16.  Those
SNe that do not converge within 16 iterations invariably have extremely
poor light-curve coverage and are excluded from the sample.  
Because high-redshift SNe very rarely have data beyond day 30, in order
to prevent a bias between high and low redshift SNe in the fitting procedure
data between 30 and 200 rest-frame days of maximum are not included,
a similar procedure to that followed in P99 and K03.
Observations more than 200 days after maximum light are included because
they provide final reference information useful for setting the amount of
host galaxy light underlying the SN.

This data set contains observations in 14 filters.
$BVRI$ filter curves were obtained from \citet{Bessell:90}.  We reiterate the
warning of \citet{Suntzeff:99} that these filter functions include a
linear function of $\lambda$, which we have removed.  The same is true
of the redshifted $B$ and $V$ filters used for some observations by the HZSST
($B35,V35,B45,V45$), with filter curves given by \citet{Schmidt:98}.
Filter curves for the {\it HST} filters on WFPC2 and ACS were generated using
{\it synphot} \citep{Simon:96synphot}.  There are two sets of ground-based 
$z$-band observations: those from \citet{Tonry:03}, and the
$z^{\prime}$ observations taken with SuprimeCam on the Subaru telescope 
presented in \citet{Barris:04}.  The \citet{Tonry:03} $Z$-band response 
curve is as presented in that paper, and the SuprimeCam $z^{\prime}$ system 
response was provided by H.\ Furusawa (2004, private communication). 

Once the date of maximum and stretch are measured, the points in the CMAGIC
linear region can be determined and the linear relation fitted.  Note that the
CMAGIC fit is performed on the observed data points, not on the template
fit used to determine the stretch and date of maximum.  Again a 
\chisq\ minimization routine is used based on MINUIT that allows for errors
in both $B$ and \BmV .  The narrowness of the CMAGIC slope distribution, as 
shown in figure~\ref{fig:slopes}, led W03 to suggest fitting all 
CMAGIC relations with a fixed slope set at the mean of this distribution.  
This is particularly important when working with high-redshift SNe because 
the observational error bars are sufficiently large that accurate slope 
measurements are difficult.  We can make better use
of the available data by assuming that low- and high-redshift SNe have
similar \betabv\ distributions, as determined by examining
low-redshift SNe.  This is similar to the approach followed by
previous analyses based on maximum magnitudes, where light-curve templates
developed from low-redshift SNe are used to fit high redshift data.
This leaves only one parameter in the fit, \Bbvzs .  However, it is possible
to test the assumption that the slope distributions are consistent 
with the handful of high-redshift SNe with sufficiently small observational 
errors (\S\ref{sec:highzdemo}).

We improve on the fixed slope assumption by numerically propagating the
additional error due to the observed distribution of slopes using
a Monte-Carlo style approach.  The slope distribution is determined from 
the low-redshift SN sample, which for this purpose includes SNe~Ia 
that are not in the Hubble flow. We take care to apply the same cuts, 
described in \S \ref{subsec:cuts}, on this sample as we
do on the sample used to directly determine the cosmological parameters,
except for the redshift cut.  This approach slightly overestimates the 
errors because the measured slope distribution includes observational 
errors, but in any case the net effect is quite small, inflating the 
errors on \Bbvzs\ by around 0.01-0.03 mag in quadrature without 
affecting the central values.  In other words, the assumption of a fixed
slope used in W03 works extremely well for current data sets,
although we do include the additional error term in this analysis.

\section{CMAGIC RELATIONS AT HIGH REDSHIFT}
\label{sec:highzdemo}
\nobreak
The first task in applying CMAGIC at high redshift is to determine if 
SNe~Ia at high redshift follow the linear relations derived at low redshift.  
A brief examination of the CMAGIC diagrams shows that high-redshift
SNe do obey linear relations between magnitude and color.  However,
in order to put this statement on a more quantitative footing,
we investigate the consistency of the \betabv\ distributions.
Most high-redshift observations have sufficiently large 
error bars that they do not provide useful slope constraints.  However, there
are a handful of relatively well observed SNe~Ia that can be used to 
investigate this question: SNe 1997ce, 1997cj, 1998aw, 1998ax, 
and 1998ba.  The requirement for membership in this set is that there
be at least three points in the CMAGIC linear region and that 
$\sigma_{\betabv} < 0.5$. SN 1997ce is particularly interesting because it
clearly displays a bump feature.  Whatever physical mechanism causes the 
bump feature is still active at high redshift.

\begin{figure}
\plotone{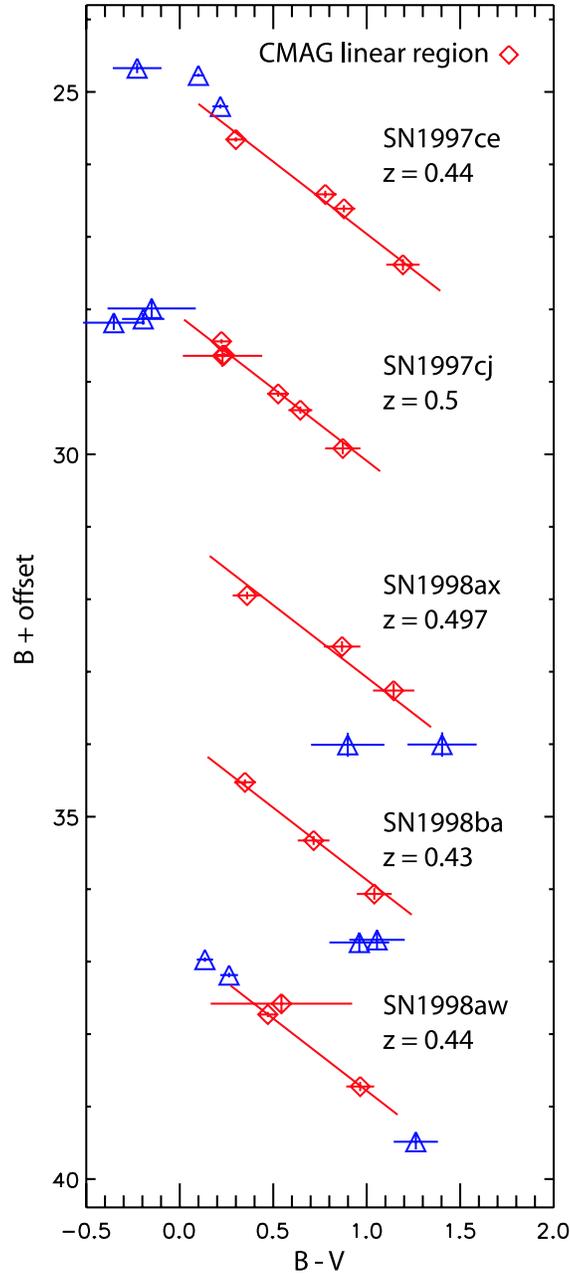}
\caption{CMAGIC diagrams for well observed high redshift
 SNe~Ia: SN1997ce (z = 0.44), SN1997cj (z = 0.5),
 SN1998ax (z = 0.497), SN1998ba (z = 0.43) and SN1998aw (z = 0.44).  
 Note the prominent bump before the linear region for SN1997ce. The 
 points in the linear regions are shown as diamonds. \label{fig:hizcmagex} }
\end{figure}

\begin{figure}
\plotone{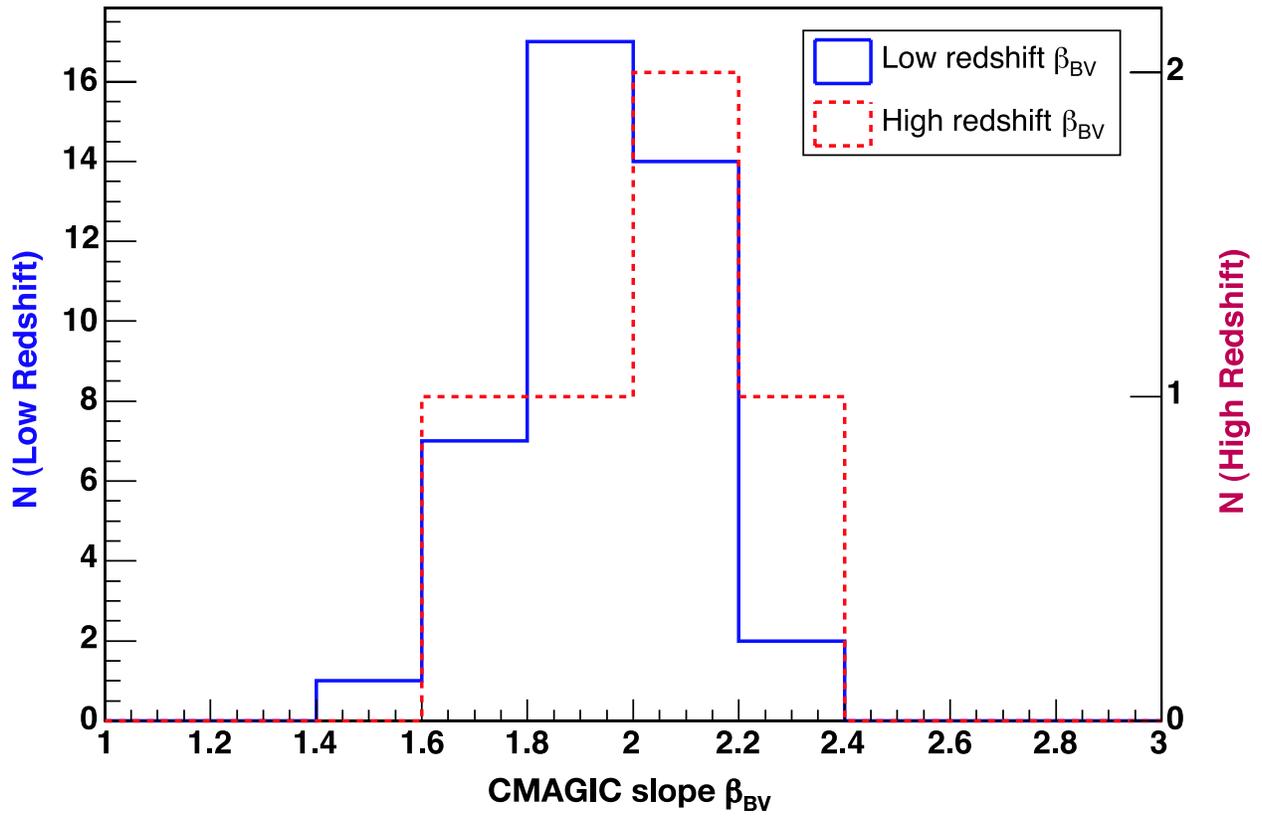}
\caption{Histogram of CMAGIC slopes for low and high redshift
 primary fit samples.  The high redshift histogram (scale on
 the right) has been scaled to the low redshift histogram
 (scale on the left) for display purposes. \label{fig:slopehisto} }
\end{figure}

The best fit slopes for these SNe are tabulated in 
table~\ref{tbl:highzslopes} and the CMAGIC diagrams are plotted
in figure~\ref{fig:hizcmagex}. The \chisq\ values for these fits
are improbably low, suggesting that the photometric errors 
have been overestimated, which is also true of the low redshift sample.  
The slopes are histogrammed in figure~\ref{fig:slopehisto}.
The mean slope for the low redshift sample is 
$\left< \betabv \right> = 1.98 \pm 0.03$ and for the high redshift sample
it is $\left< \betabv \right> = 1.96 \pm 0.11$, so there is no evidence
for disagreement.  A stronger statement requires more high quality multicolor
observations of high redshift SNe~Ia.

\section{COSMOLOGY FITTING PROCEDURES}
\label{sec:cosfits}
\nobreak
We now proceed to the primary purpose of this paper, the cosmological
analysis.  Here we describe our methodology for 
performing these fits.  The results presented here differ from previous 
papers in several respects.  First, we have attempted to formalize the 
procedure whereby individual SNe are rejected or accepted into the data 
sample to a greater extent than has been true previously.  Second, we make 
use of a blind analysis procedure in order to prevent experimenter bias
from affecting the results.  To this end, the results of the
cosmological analysis have been hidden from the authors until the cuts and 
fitting procedure were finalized. 

\subsection{Determining the Cosmological Parameters}
\nobreak
The luminosity distance equation can be written (in magnitudes) as
\begin{equation}
  m = 5 \log_{10} \left( {\cal D}_{L} \left( z, \om, \ol \right ) \right)
 + \scriptm - \alpha \left( s - 1 \right) \label{eqn:lumdist}
\end{equation}
where $m$ is the observed magnitude, $s$ is the stretch, \scriptm\ is a 
combination of the Hubble constant $H_{0}$ and the absolute magnitude of 
an SN Ia, and ${\cal D}_{L}$ is the $H_{0}$ free luminosity distance
given in \citet{Perlmutter:97}.
Because of the somewhat complicated nature of this parameter space, the
most conservative approach to fitting this relation is to perform a grid 
search over the four
fitting parameters \mbox{(\om , \ol, \al , \scriptm)} and then marginalize 
over the two nuisance parameters \mbox{(\scriptm , \al)}.  This is the 
procedure used in P99 and K03.  Because of the highly nonlinear nature
of the problem and the large errors on the cosmological parameters,
looking for the point where the \chisq\ has increased by 2.3 over its
minimum leads to an underestimate of the errors.
A \chisq\ is calculated at each
point on the grid, making use of equation~\ref{eqn:lumdist},
and converted into a relative probability 
$P \propto \exp \left( - \chisq / 2 \right )$.  The probabilities are then 
normalized over the grid, and the nuisance dimensions are summed over.  
The parameter ranges explored are $\om = [0,3]$, $\ol=[-1,4]$,
$\scriptm = [24.7, 25.5]$,\footnote{The definition of \scriptm\ used here 
differs slightly from that of P99 and K03 in that all of the constants have 
been absorbed, including c.} and $\al = [-0.5, 2.0]$. 
These ranges include more than 99.99\% of the probability.\footnote{This
could be verified prior to unblinding for \scriptm\ and \al , but
the confirmation of this statement for \om\ and \ol\ was only available
after unblinding.  If the final cosmology had disagreed very strongly
with previous results, this would have led to problems with the
blindness procedure.  Fortunately, this turned out not to be the case.}

We have also constructed fits to the equation of state parameter $w$.
In order to reduce the computational complexity of this problem, these
fits are restricted to the flat universe case.  Here the four parameters
are \om , $w$, \scriptm , and \al .  ${\cal D}_L$ must be modified 
appropriately, but in all other respects the fit procedure is identical.
The range of $w$ considered is $[0,-3.5]$.

The errors on each \Bbvzs\ include the following terms:
\begin{itemize}
\item The uncertainty from the CMAGIC fits, including a contribution from the
 distribution of \betabv .
\item The uncertainty of the stretch from the lightcurve fits multiplied by
 \al .
\item A term due to the uncertainty in redshift.  This includes an 
 assumed peculiar velocity dispersion of 300 km $s^{-1}$ and 
 redshift measurement errors .
\item \sgi\ magnitudes of intrinsic variation determined by
 fits to the low-redshift Hubble diagram. 
\end{itemize}
At high redshift the redshift measurement errors are taken to be 
0.001 when the redshift was measured from host galaxy lines and 0.01 
when measured from SN features, as in P99 and K03.
The intrinsic variation is assumed to be distributed as a Gaussian, and
is determined by performing Hubble fits with low redshift SNe and finding
the value that results in a \chisq\ per degree of freedom of 1.
A Monte-Carlo simulation was used to calculate the errors associated with
this estimate by generating 100,000 realizations of a nearby SN
sample with identical properties to the actual one (redshift distribution
and photometry errors).  For \Bbvzs\ with stretch correction,
$\sgi = 0.12^{+0.03}_{-0.04}$ mag.  Two additional estimators for
\sgi\ were considered: the RMS corrected for photometry errors and
peculiar velocities, and the maximum-likelihood (ML) estimator for this
problem.  All three agree, although the ML and \chisq\ estimators 
are considerably more efficient than the corrected RMS.  We note that this
value for \sgi\ is slightly higher than that given in W03; the values 
there were based on samples with tighter color cuts.

\subsection{Cuts on the Supernova Sample}
\label{subsec:cuts}
\nobreak
The procedure used to estimate the systematic errors in this paper
is an extension of that used by P99 and K03 and differs only in
that we have endeavored to be even more methodical in our exploration of
changes to the fits.  For this paper we specify a {\it primary} 
fit defined by a set of cuts, which are designed to be fairly loose while still
removing SNe with obviously bad data or that provide no useful constraint
on the cosmological parameters. We then explore the effects of changing these 
cuts in great detail and use the information thus gleaned to estimate the 
systematic errors.  As we discuss below, altering most of these cuts has 
little effect on the final result, but this systematic exploration raises the 
specter of an unconscious fine-tuning to obtain the
expected result.  To circumvent this possibility we have
performed a blind analysis, as detailed in \S\ref{subsec:blindness}.

The cuts can roughly be split into two categories: data quality and
analysis cuts.  Not all are used in every fit considered.
Their values for the primary fit are summarized in table~\ref{tbl:primarycuts}.
More complete descriptions are provided below.  The same cuts 
are applied when determining the sample of SNe that are
used to measure the intrinsic distribution of \betabv .  

There are four data quality cuts:
\begin{itemize}
\item A cut on the minimum number of points in the linear cmagic region.
 As long as the date of maximum is well known, it is not 
 necessary to have more than one point.\footnote{Technically a floor of 2 
 points is used when the slope distribution sample is determined, but
 this has no effect because all of the low redshift SNe have 2 or more 
 points in the linear region.}
\item A cut on the maximum allowable error on \Bbvzs .
 Objects with very poorly determined magnitudes add little statistical 
 weight to the cosmology fit but make the Hubble diagram more difficult to 
 read and in general obfuscate the result. 
\item A cut on the maximum allowable error in the date of maximum.
 This is used because the date of maximum is used to specify 
 the points that are in the linear CMAGIC region.  Points that fail this cut
 usually fail the next cut as well.
\item A cut on the maximum allowable gap (in rest frame days) 
 between the nearest point in either $B$ or $V$ and the date of $B$ maximum.  
 If this gap is too large, the date of maximum, stretch, and maximum 
 magnitude can easily be incorrect.  This arises because the error in the
 light-curve template itself is currently not fully taken into account.
\end{itemize}
There are four analysis cuts:
\begin{itemize}
\item A minimum redshift cut for the cosmology fit.
 It is ignored when the sample of SNe used to determine the intrinsic
 \betabv\ distribution is determined. 
\item A maximum redshift cutoff for the cosmology fit, which is not
  used in the primary fit.
\item A maximum allowable color excess at $B$ maximum when compared with
  the color model of K03.  This can be interpreted as an extinction cut.
\item A minimum allowable stretch value.  SNe with best fit values below
 this are removed from the sample for the reason discussed below. 
\end{itemize}

We find that our estimates for the cosmological parameters from \Bbvzs\ 
are relatively insensitive to changes in the cut on the color excess, but the 
same cannot be said of the \mb\ fits.  Because we seek to compare
the CMAGIC results directly with the \mb\ results, it is useful to
choose a value of the color cut that can be used for both fits.
Therefore, we have chosen to use the same cut as \citet{Knop:03}
($< 0.25$) in the primary fit.

A minimum stretch cut of 0.7 is applied to our primary fit sample because
our \kcorrs\ may not be reliable for extremely low stretch SNe, as their 
spectra display strong Ti features that are not well represented by our 
spectral template \citep{Nugent:02}.  We require
spectroscopic identification for our sample.  There is only one SN
that passes the other cuts but does not have a firm spectroscopic ID: 
SN 2001fo from \citet{Barris:04}.  As was the case in K03 and P99, 
SN 1997O has been manually excluded from our sample.  When 
included it is a 7 \sg\ outlier from the best fit cosmology.
Two of the low-redshift SN in 
our sample (SN 1997br and SN 1997bp) appear to have internal 
inconsistencies in their photometry, displaying a far higher degree 
of scatter both in light-curve and CMAGIC fits than can be explained 
by their quoted photometric errors.\footnote{The \chisq\ per degree
of freedom for the CMAGIC fits to SN1997bp and SN1997br are around 4, 
which is particularly striking because for the majority of SNe~Ia 
the \chisq\ per degree of 
freedom is considerably less than one.}  We have taken the conservative 
approach of removing them from the sample.  When included, they have
no impact on the cosmological parameters.  In addition to these cuts,
the maximum redshift of SNe that are used to measure the
\betabv\ distribution is specified by another cut.

There are 119 SNe at redshifts greater than 0.01 of the 131 SNe in
our baseline sample.  Lower redshift
SNe can also be included in our fits, but add essentially no statistical 
weight because of the dominance of their peculiar velocity errors.  They 
are still useful for measuring the intrinsic slope distribution.  The data 
quality cuts
at the levels of the primary fit eliminate 62 of the SNe from the primary
sample, and the analysis cuts remove five more.  53 are at $z > 0.1$, of which
28 are eliminated by the quality cuts and four by the analysis cuts.
We have explored the effects of both relaxing and tightening the cuts in
a systematic fashion.  Many of the SNe fail multiple cuts, and the cuts are 
not applied in any order, so it would be misleading to specify the number of 
SNe removed by each cut.  However, a list of which SNe are removed by
each cut is potentially interesting, and is provided in 
appendix~\ref{apndx:cutremoved}.

\subsection{Blindness}
\label{subsec:blindness}
\nobreak
 ``Experimenter bias'' occurs when an analysis is affected by the
expectations of the experimentalist.  Such bias is frequently unconscious,
and can take quite subtle forms.  For example, a result that disagrees 
strongly with a previous result is frequently subject to more 
scrutiny than one that appears to be in agreement.  This may bias an
experimenter into being more likely to find errors that cause their
result to disagree with expectations while making it less likely
that they will discover errors that have the opposite effect.  Since the
research process has a natural termination point (publication), if the
decision to stop analyzing a result is at all influenced by the value
of the result, a bias will be introduced.  A nice
summary of these issues can be found in \citet{Heinrich:03}.  It has
long been recognized that a useful technique for mitigating experimenter 
bias is to hide the final results of the experiment 
from the experimenter for as long as possible. This is known as blind analysis.
Such an approach is particularly useful in an analysis with a substantial
number of cuts, such as that presented here.
In the medical fields double blind procedures (which hide some details
of the experiment from both the test subject and the experimenters)
are used almost as a matter of course.  Naturally,
hiding the details of the experiment from the subject is not of great
concern in astronomical research.  

A critical point is that these techniques do not seek to completely
hide all information during the analysis.  In fact, the goal is to hide
as little information as possible while still acting against experimenter
bias.  Human judgment and scientific experience continue to play a
critical role in a blind analysis.  One does not mechanically carry out
the steps of the analysis and then publish the results.  All that a
blind analysis does is prevent unconscious misuse of particular types
of information during the analysis process.  The kind of data that are
excluded from consideration (namely, the final answer derived from each
option under consideration) is invariably that which no reasonable 
scientist would allow to consciously influence his or her decision making 
process.  However, subconscious effects are still present, and this
is what this approach helps prevent. 

Specifically, it is important to design the blindness technique such that
subsidiary diagnostics are available even while hiding the
final answer.  Errors are initially present in any analysis, and it is 
important that even while the result remains blinded mechanisms are available 
to catch these problems.  Specifically, our goal is to hide the values of \om\ 
and \ol\ until the cuts and fitting procedures have been finalized, while
preserving as much ancillary information as possible.  In particular, our 
method preserves the residuals of individual SNe with respect to the 
Hubble line, which is extremely useful while diagnosing the fits.  For example,
an error in the \kcorrs\ might result in all SNe in a given redshift range 
departing significantly from the Hubble line.  This problem would still be 
detectable in our blinded fits.  In addition, the method preserves the 
shifts in \om , \ol\ between fits to different subsamples -- if excluding a 
particular SN causes 
the unblinded result to shift by $\Delta \om = 0.1,\ \Delta \ol = 0.2$, the 
blinded result shifts by the same amount, which is important when investigating
systematic errors. 

The technique used here is based on altering the true fit estimates.  Hidden,
but fixed, offsets are added to \om\ and \ol , and this change is propagated
through to the \Bbvzs\ values.  In essence the cosmological parameters
are fitted twice, with the magnitudes modified between fits, but the
results of the first fit are never output.  Because it would be possible
to circumvent the blindness if the real \Bbvzs\ values were known,
these values must be kept hidden.  All of the programs used to
plot CMAGIC diagrams add random offsets to the $B$ magnitudes for display 
purposes. Furthermore, the CMAGIC fitter and cosmology fitter are integrated 
so that the true \Bbvzs\ values are not output.

The expression for the luminosity distance cannot 
be evaluated in terms of simple functions except in limited cases, 
so the magnitude modification is calculated numerically.  The results of the 
first, unmodified, fit are marginalized to determine the secret true measured 
values $\Omega_{mT}$ and $\Omega_{\Lambda T}$.  The 
hidden offsets are then applied to these values, and the difference in 
magnitudes between the two cosmologies is calculated and applied.
If $\Delta \om$ and $\Delta \ol$ are the hidden offsets, 
then the following function is added to \Bbvzs\ for each SN:
\begin{equation}
 \Delta \Bbvzs \left( z \right) = 5 \log_{10} 
  {\cal D}_{L} \left( z, \Omega_{mT} + \Delta \om ,
                  \Omega_{\Lambda T} + \Delta \ol \right) -
  5 \log_{10} {\cal D}_{L} \left( z, \Omega_{mT}, \Omega_{\Lambda T} \right),
 \label{eqn:magshift}
\end{equation}
where ${\cal D}_{L}$ is as in equation~\ref{eqn:lumdist}.
The cosmological fit is then redone with the new magnitudes and this result
is output.  It is safe to output the modified magnitudes, which can be used
to construct a Hubble diagram and to perform various tests on the fit.

The simplest method to choose the hidden offsets is to generate them randomly.
This performs poorly in this case because there are several non-physical 
regions in the \om , \ol\ parameter space.  Negative values of \om\ result 
in a non-convergent luminosity distance integral.  For high values of \ol\  
the universe did not experience a Big Bang, but is instead rebounding from a 
previous bout of contraction \citep{Carroll:92}.  In such a universe 
there is a maximum observable redshift, and if any of the SNe are at higher 
redshifts the luminosity
distance expression cannot be evaluated.  A randomly generated
offset could easily push the cosmological parameters into one of these
regions.  Instead we have chosen to generate the hidden offsets by specifying
the desired values of \om\ and \ol\ for a particular SN sample 
(the primary fit).  A special version of the cosmological fitter determines 
the offsets between a fit to the primary sample and the chosen 
value\footnote{These values were chosen to be sufficiently different from the
results of previous analyses to force internal reviewers to psychologically
confront the blindness scheme while remaining close enough to the expected
values that the resulting error contours were not overly distorted.}  \om = 1, 
\ol = 1.1.  These offsets are then used for all other fits.

As long as the resulting fit values for \om\ and \ol\ are roughly 
equal to ($\Omega_{mT} + \Delta \om$, $\Omega_{\Lambda T} + \Delta \ol$)
this preserves the residuals with respect to the fit by construction.  
Because the same hidden offsets are used
for all fits, this approximately preserves relative shifts between different
fits.  The caveat is that, for a particular value of \om\ and \ol , the
shape of the luminosity distance equation effectively weights SNe
depending on their redshift, and therefore altering the values of these
parameters may cause the relative shifts in the blinded fits to be slightly
different than for the true values.  Therefore, the offsets are determined
iteratively.  However, as long as the hidden offset
is relatively small, this effect is negligible.  Tests on both previous
data sets (specifically, the low-extinction primary subset of K03) and 
artificially generated data show that this procedure works in that
the resulting cosmological parameter estimates are equal to the unblinded
result plus the specified offset.  The offset between the blind target
values and the actual estimates for this analysis was somewhat larger
than anticipated, so the specified offset does not quite match the
actual shift.  However, the relative shifts are preserved accurately over
small distances, which allowed us to compare different fits to the same
data prior to unblinding.

A similar procedure is followed in the $w$ fits, although a different set of
offsets are used.  Because problems related to non-physical regions
of the parameter space are not as severe in this case, the offsets to
\om\ and $w$ were randomly generated from the ranges $[-0.2,0.2]$
and $[-0.4,0.4]$.

Should a mistake in the analysis be found after the result is unblinded,
it should still be corrected.  In this situation, one should
publish both the corrected and uncorrected results and note the effects
of the discovered error on the result. An example of this can be found in
\citet{Akerib:04}. We also note that it is important to determine the
systematic errors prior to unblinding, or it would be possible to explain 
away any unexpected results by inflating them.  This technique certainly 
does not prevent all types of bias, but it does provide an 
opportunity to improve the situation, and thus is worth pursuing.

\subsection{Complete Fitting Procedure (Blind)}
\nobreak
Our cosmological fits proceed in the following order:
\begin{itemize}
\item The SNe used to measure the intrinsic \betabv\ distribution
 are determined by applying the specified cuts.  The distribution of \betabv\ 
 is then calculated from these SNe.
\item A one-parameter (\Bbvzs ) CMAGIC fit is performed for all SNe in the data
 sample using a Monte-Carlo fitting technique that takes into account the 
 distribution of \betabv\ from the distribution calculated in the previous 
 step.  The fitted \Bbvzs\ values are not output.
\item The cuts are applied again to determine the SNe used to
 measure \om\ and \ol .  The same cuts are used,
 except for the redshift ranges in \S \ref{subsec:cuts}.
\item A cosmological fit is performed.  Estimates for \om\ and
 \ol\ are calculated but not output.
\item The hidden offsets are read in and added to \om\ and \ol .  A
 magnitude offset is applied to each SN based on equation~\ref{eqn:magshift}.
\item The cosmology is refitted with the new magnitudes.  These results
 are output.
\item The altered magnitudes are used to construct a Hubble diagram.
\end{itemize}
Once the blindness was removed, the fits were redone without the secret 
offset step.  We have also performed fits using the maximum $B$ magnitude, 
\mb .  Since these fits are not a principal result of this paper they can be 
performed in an unblinded fashion, allowing us to test our procedures.  

\section{COSMOLOGICAL RESULTS}
\label{sec:cosresults}
\begin{figure}
\plotone{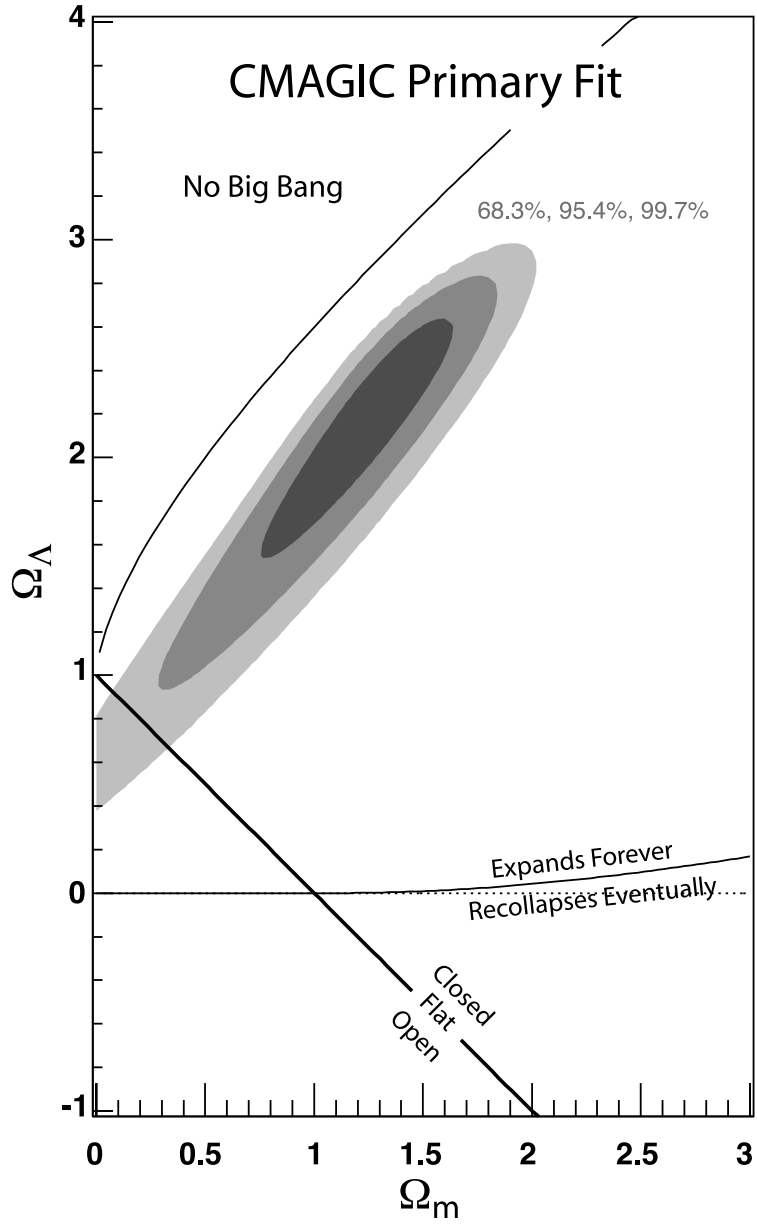}
\caption{Confidence regions for \om , \ol\ from the primary fit.  The
contours represent 68.3\%, 95.4\%, and 99.7\% of the total probability.
 \label{fig:baselinecontour}}
\end{figure}

\begin{figure}
\plotone{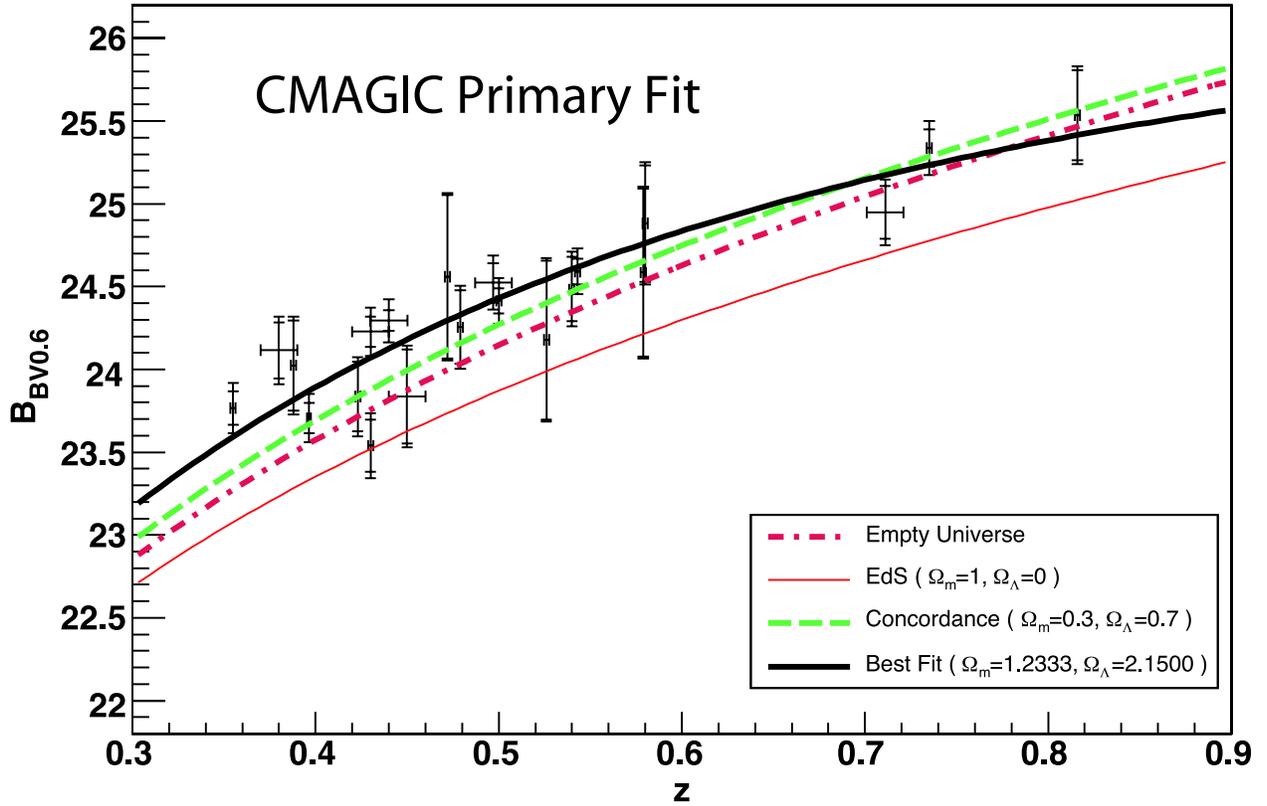}
\caption{Hubble diagram for the primary fit.  The outer error
bars include both $\sgi = 0.12$ of intrinsic scatter derived from low
redshift fits and the contribution of the peculiar velocity errors
(which are also shown as the horizontal error bars); the inner ones do not.
Also shown are the relations for an empty Universe, an Einstein-deSitter
one, and the ``concordance'' cosmology with $\om = 0.3$, $\ol = 0.7$.
\label{fig:baselinehubble}}
\end{figure}

\nobreak
Figure~\ref{fig:baselinecontour} shows the \om , \ol\ confidence regions of
our primary fit, based on 31 nearby and 21 distant SNe~Ia.  An additional nine 
very nearby SNe ($z < 0.01$) are used while determining the \betabv\ 
distribution (for a total of 40).  The resulting estimates for
the cosmological parameters are
$\om = 1.26^{+0.38}_{-0.51}$ and $\ol = 2.20^{+0.41}_{-0.67}$.  If we 
require a flat universe, consistent with recent CMB results, then 
$\om = 0.19^{+0.06}_{-0.06}$.  These confidence regions are comparable to 
those from P99 (but not as good as those from K03), despite the fact that 
fewer SNe are involved, due to the smaller value of \sgi\ for CMAGIC.  
The fit residuals are shown in figure~\ref{fig:baselinehubble}.  

\begin{figure}
\plotone{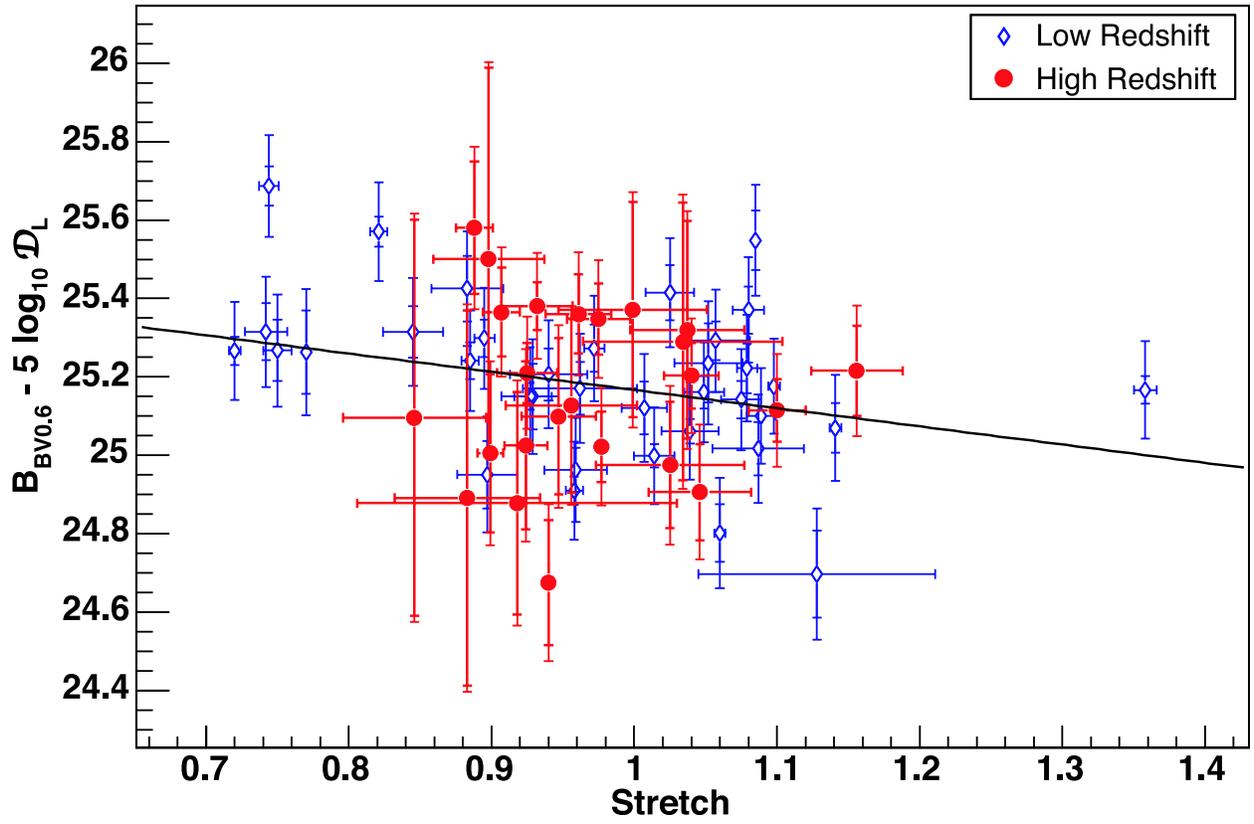}
\caption{Stretch-luminosity relationship for the low redshift SNe
 (open diamonds) and high redshift SNe (filled circles).  Each point
 is \Bbvzs\ minus the $c/H_{0}$ free luminosity distance of
 equation~\ref{eqn:lumdist}, plotted against the stretch of the SN.  The
 line represents the fit estimates for \al\ and \scriptm\ from the
 primary fit (\al = 0.52, \scriptm = 25.17).  The outer error bars
 include 0.12 magnitudes of intrinsic scatter.\label{fig:stretchlum} }
\end{figure}

\om\ and \ol\ are not the natural variables for this measurement, as they 
are not independent for this data set.  The result of our analysis is 
better expressed in the principal axes frame of the error ellipse
$\Omega_1 \equiv 0.790 \om - 0.613 \ol$ (the short axis) and
$\Omega_2 \equiv 0.613 \om + 0.790 \ol$ (the long axis).
Roughly, $\Omega_1$ can be thought of as measuring acceleration and
$\Omega_2$ as measuring geometry.  Analyzing the results in this frame
has considerable benefits while calculating systematic errors and when
comparing the CMAGIC results to those derived from maximum magnitudes.
In this frame the results of the primary fit are 
$\Omega_1 = -0.349^{+0.117}_{-0.131}$ and 
$\Omega_2 = 2.502^{+0.530}_{-0.838}$. 
The values of the nuisance parameters are
$\al = 0.516^{+0.193}_{-0.206}$ and $\scriptm = 25.166^{+0.049}_{-0.045}$,
and they are almost completely statistically independent.
Magnitudes and redshifts are provided in table~\ref{tbl:primarysamplowz} 
for the low-redshift sample, and in table~\ref{tbl:primarysamphighz} for the
high redshift sample.  The \chisq\ of this fit is 49.5 for 
52 degrees of freedom.  In the next section we discuss variations of the 
cuts, which produce different sets of SNe.    The stretch-luminosity
relation is shown in figure~\ref{fig:stretchlum}.  When compared
with the \mb\ relation (Fig.\ 13 of \citet{Knop:03}, for example), the 
evidence for the utility of a stretch correction is much weaker for \Bbvzs .  

\begin{figure}
\plotone{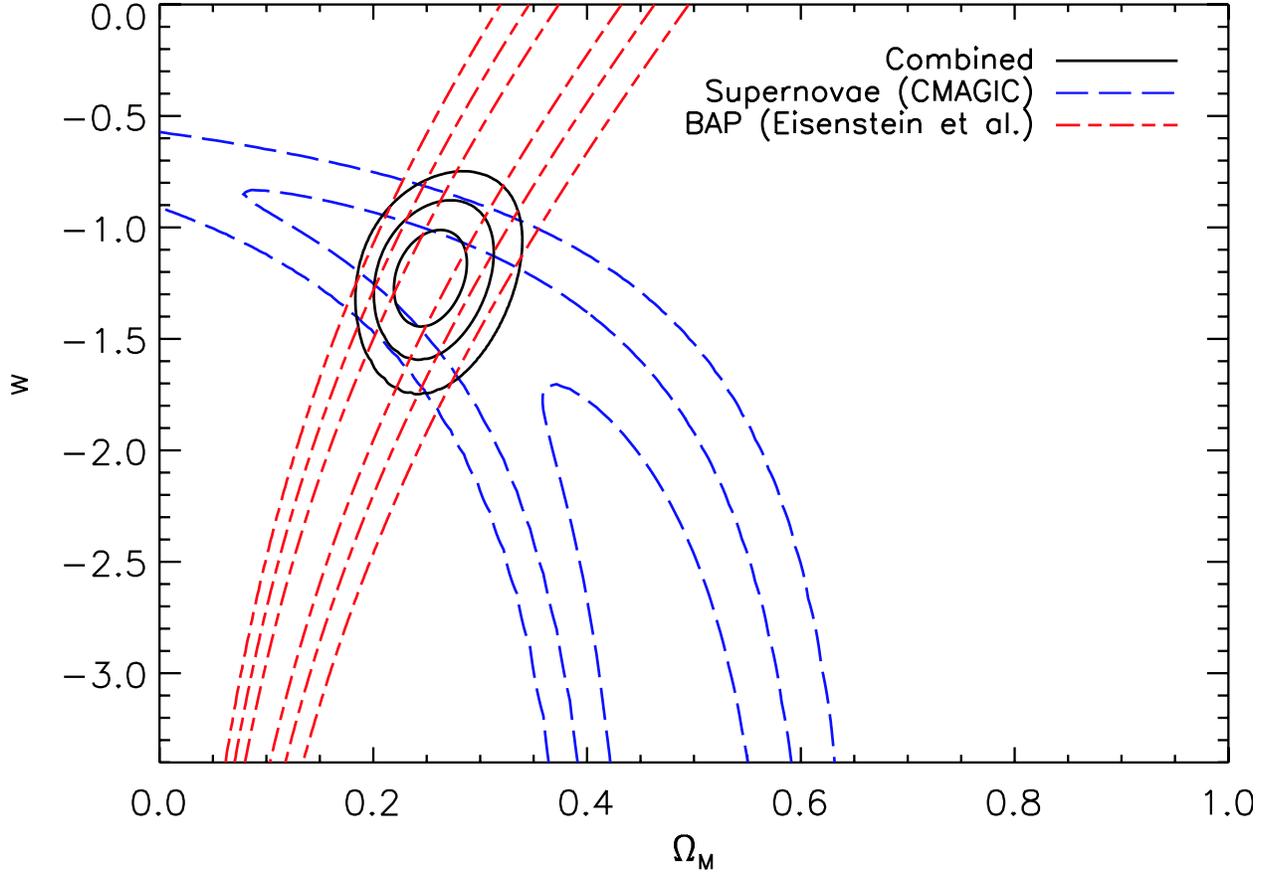}
\caption{Confidence regions for w, \om\ assuming a flat universe.
 The contours represent 68.3\%, 95.4\%, and 99.7\% of the total
 probability.  Both the constraints from the CMAGIC analysis
 of supernovae and the baryon acoustic peak (BAP) of
 \citet{Eisenstein:05} are shown, as are the contours that
 result from combining the two measurements.
 A cosmological constant corresponds to $w = -1$. \label{fig:wcontour}}
\end{figure}

Our estimates for $w$ in a flat universe are shown in 
figure~\ref{fig:wcontour}.  These are combined with the measurement of 
the angular size of the baryon acoustic peak (BAP) in SDSS galaxy clustering 
statistics at $z=0.35$ \citep{Eisenstein:05}, which are quite complementary 
to the SN measurements.  The resulting constraint is
$w = -1.21^{+0.15}_{-0.12}$ and $\om = 0.25^{+0.02}_{-0.02}$ (statistical 
errors only).

\begin{figure}
\plotone{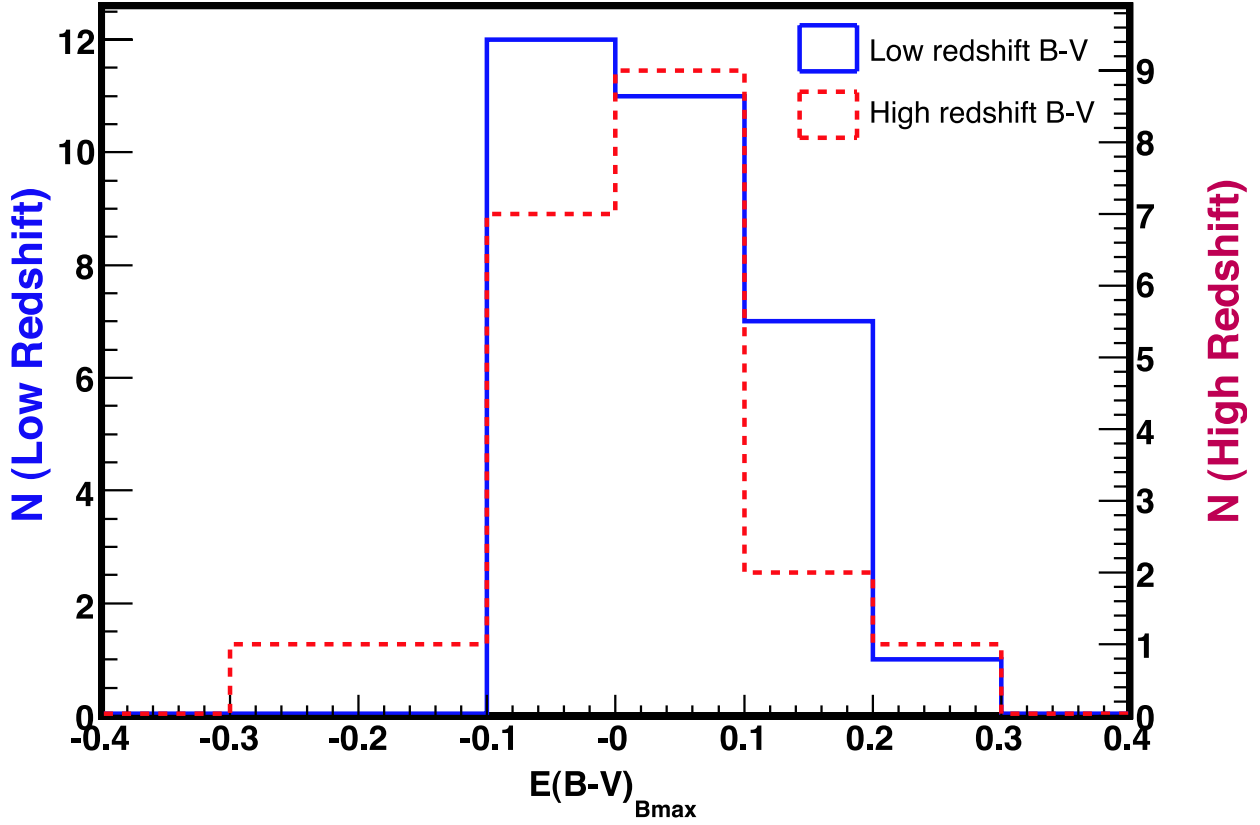}
\caption{Histogram of \EBmVBmax\ at $B$ maximum for the low and high redshift
primary fit sample.  The high redshift histogram (dashed line,scale on right) 
has been scaled to the low redshift histogram (solid line, scale on left) for 
display purposes. The mean color of the low redshift sample is 
$0.045 \pm 0.027$ and that of the high redshift sample is 
$0.027 \pm 0.019$. The two distributions are consistent given the small
number of events in each bin.\label{fig:maxcolor}}
\end{figure}

This is the first analysis that treats
the combined data from the different SN groups in a fully consistent manner.
Unlike \citet{Leibundgut:01} or \citet{Riess:04}, we find no significant
evidence for anomalously blue colors in the high-redshift SN, even though 
this sample contains many of the same objects as those studies.  
Figure~\ref{fig:maxcolor} shows the \BmV\ color at $B$ maximum for the 
low- and high-redshift primary fit sample.  The highly negative color 
point from the high-redshift sample is due to (by far) the most poorly 
measured SN, SN 1997af, which has $\EBmVBmax = -0.24 \pm 0.24$.  Excluding 
this point, the mean color of the low redshift sample is 
$\BmVBmax = 0.045 \pm 0.027$ and that of the high redshift sample is 
$\BmVBmax = 0.027 \pm 0.019$, where the standard errors are quoted.

\section{SYSTEMATICS}
\label{sec:systematics}
\nobreak
We explore various systematic errors by performing alternate fits and 
comparing the results with our primary fit.  Because of the way in which our
blindness scheme is constructed, this comparison was possible before the 
final answer was known.  As was the case in \cite{Knop:03}, we find that 
the effects of most of the systematics act along the long axis of our error 
ellipse. They therefore do not significantly affect the value of the SN
measurements for determining if the Universe is accelerating, but do
substantially limit our ability to measure geometry.  Fortunately, this is 
the dimension in which CMB measurements are extremely powerful. 

There are two types of systematic error possible in this analysis.
First, there are the systematics arising from alterations in the fitting
procedures, \kcorrs , etc.  Second, there are those arising from the
cuts applied to the sample.  Ideally this second set would be handled
by a complete Monte-Carlo simulation of the SN sample.  
Unfortunately, there are far too many pieces of information missing to 
make the results of such a study at all useful.  In order to construct 
a believable Monte-Carlo, it would be necessary to have a reasonable 
understanding of the intrinsic luminosity and extinction distributions, 
which have not been convincingly measured.  To make matters substantially 
worse, it would also be necessary to have a good understanding of the search 
and follow up strategy used to construct the SN sample.  Because the 
sample used in this paper is primarily constituted of literature SN, 
a clear definition of the search techniques and procedures is simply not 
available.  Providing the results of such a procedure would provide a 
misleading sense of accuracy.  We therefore proceed by 
calculating the effects of changing the cuts applied to our sample over
what we consider to be a reasonable range and combining the resulting
shifts as an estimate of the systematic error.  Clearly this procedure
is somewhat subjective, but any credible improvement requires the availability
of large, well defined SN samples such as those that should be
provided by the SNfactory, SNLS, SDSS Supernova Survey, and ESSENCE.

\begin{figure}
\plotone{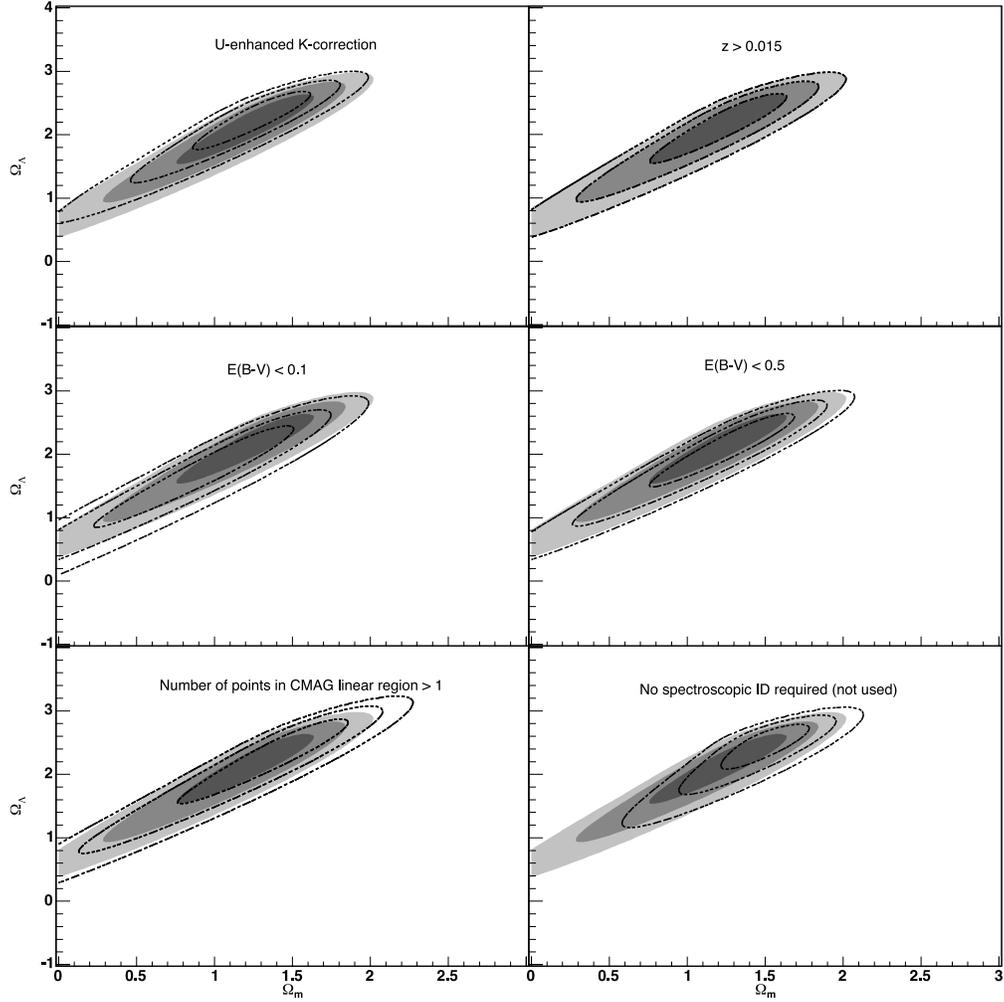}
\caption[Examples of systematics considered]
{Examples of some of the systematics considered in this analysis.  
 More complete descriptions can be found in the text.  In each panel, 
 the primary fit is shown as filled contours and the fit with the specified
 change is shown as dashed contours.  For both sets the contours
 correspond to 68.3\%, 95.4\%, and 99.7\% of the total probability.
 Note that the scenario represented in the bottom right panel (not requiring
 spectroscopic ID) is not used in the final systematics estimate,
 as this is considered unmotivated. \label{fig:baseline_comp} }
\end{figure}

The effects of these shifts can most precisely be stated in terms of
the principal axes of the primary fit error ellipse, $\Omega_1$
and $\Omega_2$, which is the primary justification for their
use.  Recall that for the primary fit $\Omega_1 = -0.349^{+0.117}_{-0.131}$ 
and $\Omega_2 = 2.502^{+0.530}_{-0.838}$ (statistical errors only).
We follow the standard practice of adding the negative and positive
shifts in quadrature when handling asymmetric errors (however,
see \cite{Barlow:03} for criticism of this procedure). The resulting 
systematic errors are $^{+0.060}_{-0.062}$ on $\Omega_1$ (the short axis), 
$^{+0.476}_{-0.545}$ on $\Omega_2$ (the long axis), and 
$^{+0.029}_{-0.049}$ on the value of \om\ in a flat universe. The shifts
are summarized in table~\ref{tbl:identifiedsystematics}, and detailed
individually in the following sections.  Some representative examples
can be seen in figure~\ref{fig:baseline_comp}.  An essentially identical
procedure has been carried out for the fit to $w$, \om\ in a
flat Universe, including the BAP constraint, resulting in systematics 
error estimates of $^{+0.07}_{-0.12}$ on $w$ and $^{+0.01}_{-0.01}$ on \om .  
Note that this only includes the systematics from the SN measurement.
Unlike the \om , \ol\ fits, here the statistical errors are dominant, 
reflecting the more challenging nature of the $w$ measurement.

\subsection{Variation of Fitting Procedures}
\label{subsec:systematicsfitting}
\nobreak
There are many reasonable ways to alter the CMAGIC fitting procedures
that result in slightly different values of the cosmological parameters.
We have attempted to explore some of these variations.

P99 found that using a floating value of \al\ when propagating
the stretch error into the fit magnitude artificially inflates \al ,
as this decreases the \chisq\ by increasing the magnitude errors.
Therefore, \al\ was fixed for the purposes of error propagation.
As in K03, we find no evidence for this effect.  Fixing \al\ at the estimate 
from the primary fit ($\al = 0.5$) has essentially no effect on the 
\om , \ol\ values except to shrink the error bars slightly, as expected.  
Not performing a stretch correction ($\al = 0$) shifts the error ellipse 
primarily along $\Omega_2$ by 0.06.  This is not included in the final 
value for the systematic error.

It is possible to include estimates about the error in the stretch
and date of maximum in the CMAGIC fitting procedure, since they influence
which points are included in the CMAGIC fit.  A modified version of
the fitting code has been used to investigate this possibility.
This approach is considerably more expensive computationally, and for
this data sample it turns out to make no difference.  In our fits we
have effectively assumed that $B$ and \BmV\ are independent variables.
An alternative formulation of the linear relations that treats $B$ and $V$
as independent variables is possible. This also has no 
effect on the fit values (less than 0.005 mag for any SN).  

The light-curve fitting procedure used in P99 differs slightly from that
used here (and by K03) in that the fits to the $V$ band were performed fixing
the stretch and date of maximum to the values derived from a $B$ only
fit.  This procedure arose from concerns that the rest frame $V$
light curves for the high-redshift sample are more poorly sampled
than the rest frame $B$ light curves, which is not the case for the 
low-redshift sample.  Thus, a light-curve fitting procedure that treats both
bands on an equal footing might effectively introduce a bias in the
fits.  This is of considerably less concern for this data sample, since
by its nature CMAGIC demands good $V$-band coverage, but to guard
against this problem we re-calculated all of the lightcurve fits following
this prescription, which affects the CMAGIC fits because it changes the
values of the stretch and date of maximum.  The resulting effect on the
error contours was minor, and primarily towards larger values
of $\Omega_2$ by 0.144.

Variations in the \kcorrs\ are investigated by considering alternative
versions of the spectral template.  In particular, we follow K03 by making use
of a $U$-enhanced version of the template with $U - B = -0.5$ instead
of $-0.4$ as in our primary fit.  This shifts the error ellipse primarily
along the short axis, with 
$\Delta \Omega_1 = -0.052$ (towards smaller values of \om ).
The \chisq\ worsens slightly to 50.9.  This is, by far, the
most significant source of uncertainty related to alterations in the
fitting procedures.  Simply treating this error as a statistical contribution
to each SN is a completely inadequate representation of its effect
on the cosmological results.  Clearly, future projects would benefit
substantially from additional constraints on the $U$-band behavior of
SNe~Ia.

\subsection{Variation of Cuts}
\nobreak
We considered both increasing and decreasing the cut values for
all of the cuts described in \S \ref{subsec:cuts}.  Here we only
present those that had a measurable effect on the error ellipse
or are interesting for some other reason.

Requiring SNe to have observations within 5 rest-frame days of maximum
eliminates two low redshift SNe (SN 1998ab and SN 2000fa)
and one at high redshift (SN 1996E), and induces a shift along the
long axis by $\Delta \Omega_2 = +0.139$.  Loosening the requirement
to 10 days adds one high-redshift SN (SN 2001jp), and results in a shift
along the $\Omega_1$ axis of +0.024 towards higher values of \om .
Changing the minimum allowable redshift to 0.015 from 0.01 has an
extremely small effect on the fit results while eliminating six
low redshift SNe.  Halving (to 0.25) or tripling (to 1.5) the cut on the 
maximum allowable magnitude error alternately removes five high 
redshift SNe or adds one, but does not affect the results substantially, 
as one would expect given the low weight given SNe with such 
large errors.

Placing a substantially tighter cut on the color at maximum
[$\EBmVBmax \le 0.1$, similar to that used for the low-extinction 
subset of K03] shifts the error contours by
a substantial amount along the long axis (towards a flat universe)
by $\Delta \Omega_2 = -0.467$, eliminating three high
and eight low redshift SNe. Using a color cut of 0.125 (half of the primary
fit value) is not substantially different than using 0.1.
Relaxing the color cut to 0.5 adds two high-redshift
(SN 1998aw and SN 2002ad) and four low redshift SNe, and 
moves the contours principally along
the short axis by $\Delta \Omega_1 = -0.048$.
While less affected by extinction than \mb , CMAGIC is not
completely unaffected.  The analysis presented in this paper suggests
that assumptions about the extinction law are not a significant systematic
bias, and therefore future studies, including those that use CMAGIC,
may benefit by applying an extinction correction. This must
be weighed against the decrease in independence of the two magnitudes
after correction.

Requiring that the date of maximum be known
to better than 0.5 days removes a large number of high redshift
SNe from the sample (nine), but has little effect except to inflate
the error contours along the long axis.
Relaxing the requirement to 2 days adds eight
poorly measured high-redshift SNe and shifts the ellipse outwards
along the long axis by $\Delta \Omega_2 = +0.115$.

Requiring that there be at least two observations in the CMAGIC
linear region, and hence providing some level of confidence that
the CMAGIC relations are being obeyed, does have a non-negligible 
effect on the cosmological parameters.  Three high-redshift 
SNe are eliminated (SN 1998as, SN 2002ab, and
SN 2002kd), and the error ellipse shifts primarily outward
along the long axis by $\Delta \Omega_2 = +0.23$. Even when two points are
required in the linear region, the quality of the high redshift data 
is such that the CMAGIC slope \betabv\ cannot be usefully fitted to
each SN.

As can be seen from the above discussion, the primary systematic
effect related to the cuts on the SN sample is associated
with the extinction cut.  A better understanding of the extinction
distribution would help reduce this systematic considerably.
Note that we do not apply an extinction correction,
so we are more sensitive to the extinction cut than some other analyses --
although they trade this off with sensitivity to extinction and the
intrinsic peak color of SNe~Ia.  Fortunately, the systematics arising from the 
cut selection are primarily along the long axis of the error ellipse,
and hence have little effect on our detection of acceleration.

\subsection{Other Systematics}
\label{subsec:othersys}
\nobreak
We have also considered limiting our low-redshift SN sample to only those
from large, systematic SN studies in order to limit any systematic 
errors arising from differences in calibration.  There are three major 
low-redshift samples: \citet{Hamuy:96}, \citet{Riess:99a} and 
\citet{Jha:05}.  Excluding all nearby SNe that are not from one of the 
above three sources has a very minor effect.

To test the sensitivity of our results to individual SNe, we have performed
a jack-knife test by removing each of the 21 high-redshift SNe individually
and recalculating the cosmological fit.  Our values for \om\ and \ol\ are
sensitive to the removal of SN 2001ix and SN 2002kd, both at the very high 
redshift end of the sample.  Removing either of these SNe shifts the
contours primarily along the long axis, although in opposite senses.
Removing SN 2001ix results in a shift inward of $\Delta \Omega_2 = -0.28$,
and removing SN 2002kd shifts the contour outward by $\Delta \Omega_2 = 0.31$.
Interestingly, their effects on the cosmological parameters nearly cancel.
This analysis would benefit from additional SNe in
this redshift range, but overall the results are reasonably robust.

Properly speaking, \sgi\ should be another quantity that is marginalized 
over while performing the cosmological fits.  To determine if this is
necessary, we performed fits in which \sgi\ was varied by 1 \sg\ in each
direction, and found that the effects on the cosmological parameters
were negligible (less than 0.1 \sg\ in $\Omega_1$ and $\Omega_2$).

Since all of the high-redshift supernovae (and many of those at
low redshift) come from flux-limited samples, they suffer
from Malmquist bias \citep{Malmquist:36}.  We note that only
a difference in the amount of Malmquist bias between the low- and
high-redshift SN samples can affect the cosmological results.
This effect is discussed extensively in P99 and K03, and we adopt 
the estimates contained therein for these samples: 0.01 mag for P99 and 
0.03 mag for K03.  P99 also estimated the Malmquist bias for the
\citet{Hamuy:96} sample as 0.04 mag.  The \citet{Riess:99a} and
\citet{Jha:05} samples were primarily discovered using a 
galaxy catalog search, so they may suffer from little or no Malmquist
bias \citep{Li:01}.  We therefore adopt a Malmquist bias of 0 mag 
for these samples.  It is difficult to estimate the Malmquist bias
for the remaining SNe in the low redshift sample, since they were
discovered in a rather inhomogeneous fashion.  However, since they constitute
only a small fraction of the sample, the effects of any Malmquist bias
on the cosmological parameters from this sample are expected to be negligible,
and so we adopt a value of 0 mag.  For the remaining portion of
the high-redshift sample (approximately half) we provisionally use
the same value as for the P99 SNe, 0.01 mag. To test the effects of
this bias on our estimate, we apply the offsets to each sample and
recalculate the fit.  The resulting shift in the cosmological parameters is
quite small, less than $0.1\ \sigma$ in both dimensions.

Appendix~\ref{apndx:bumps} contains a discussion of the effects
of the `bump' in the CMAGIC diagram exhibited by some SNe.  The effects
of this systematic are negligible along both axes (less than 0.05 \sg ).

\section{ANALYSIS OF RESULTS}
\label{sec:analysis}
\nobreak
There are two channels available for analyzing the results of
this paper.  First, the estimates of the cosmological parameters
can be considered in isolation.  Second, the CMAGIC results can
be compared with a maximum magnitude fit to the same SN.
Several of the systematics should affect both samples equally
(e.g., Malmquist bias); therefore, this comparison should
be more precise.  However, this requires that the covariance
between \mb\ and \Bbvzs\ be determined.

\subsection{Constraints on the Cosmological Parameters}
\nobreak
The results of a CMAGIC fit to currently published SN data
strongly favor an accelerating Universe --- in fact, more strongly
than previous results based on \mb .  Perhaps more interesting is that 
the fit contours depart mildly from a flat universe.
In the principal axis frame, a flat universe corresponds
to $\Omega_2 = 0.756 \pm 0.010$ for $\om = 0.191$. Once systematics
are taken into account, the disagreement is 1.75\sg , which 
is expected to occur approximately 8\% of the time
due to random chance.  A similar result was seen in the SN sample of
\cite{Tonry:03}, although at a somewhat lower level of significance.
Both results are interesting, but not yet strong enough to be of
serious concern.  One of the lessons of blind analyses is that
1.5+\sg\ disagreements occur in science more frequently
than our intuition, developed from exposure to non-blind experiments,
often expects.\footnote{See \citet{Heinrich:03} \S4 for further discussion.}

The departure from flatness is driven by SNe at moderate redshifts
$0.3 < z < 0.5$.  The three with the highest pull are
SNe 1998as, 1996k, and 1997ce.  It is difficult to find any common
thread between them.  They come from three different papers, were
observed with different telescopes (although SN 1998as and SN 1997ce
were both partially observed with {\it HST}), and their photometry 
was reduced by
different authors using different techniques.  Since they constitute
the low-redshift end of their respective surveys, there may be a suspicion 
that they suffer from unusually high extinction.
While SN 1998as does suffer from considerable host galaxy extinction
($A_V = 0.49$; K03), the other two suffer from negligible extinction
($A_V=0.02$ and 0.08 for SN 1996K and SN 1997ce, respectively;
Riess \etal\ 2004).
Note that removing each of these SNe individually has little effect on our
results, as explained in \S \ref{subsec:othersys}.

\subsection{Comparison of \Bbvzs\ and \mb\ Results}
\label{subsec:compare}
\begin{figure}
\plotone{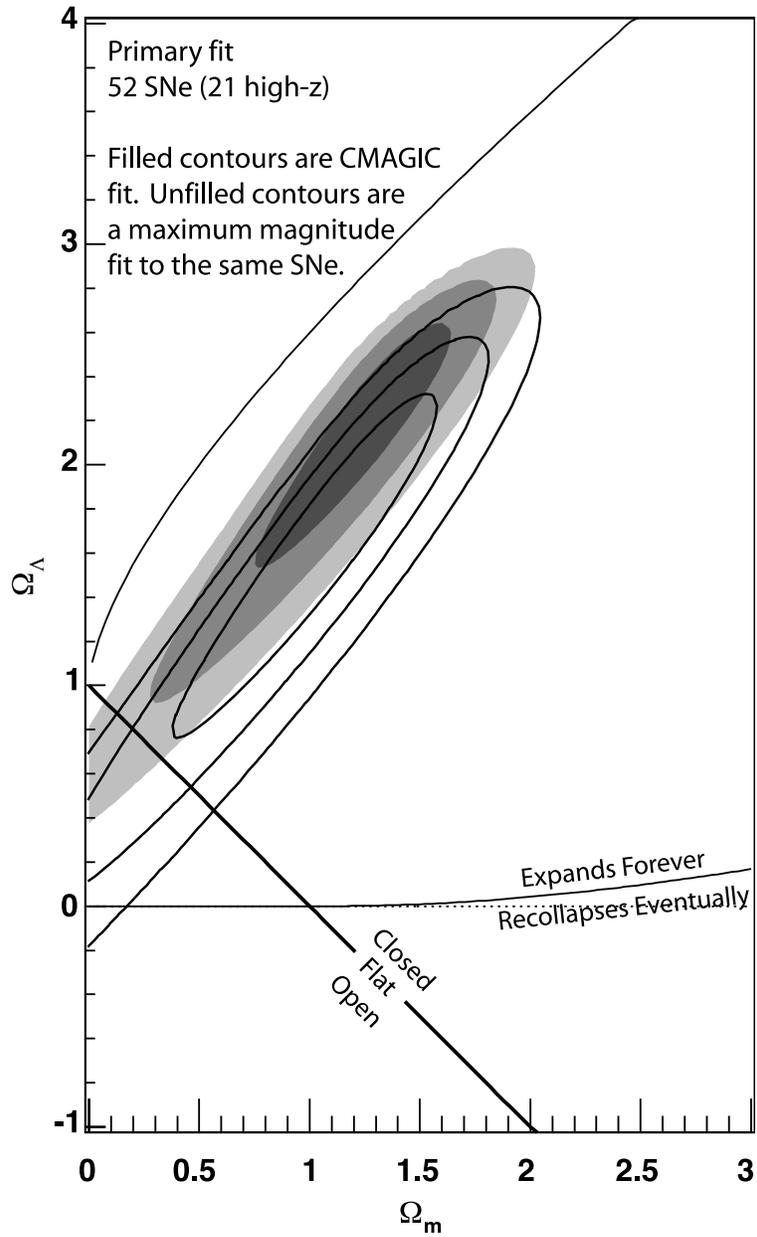}
\caption[Direct comparison of \mb\ and \Bbvzs\ results]
{ Direct comparison of the \mb\ and \Bbvzs\ cosmology
results.  The \mb\ results are shown as unfilled contours
and the \Bbvzs\ results as filled ones.  In both cases, the
contours represent 68.3\%, 95.4\%, and 99.7\% of the total 
probability. \label{fig:contcompare} }
\end{figure}

\nobreak
The results of an \mb\ fit to the same SN as the primary
are compared with the \Bbvzs\ fit in figure~\ref{fig:contcompare}.
The \chisq\ of this fit is 44.32 for 52 degrees of freedom,
and the resulting estimates are $\om = 1.08^{+0.49}_{-0.69}$ and 
$\ol = 1.65^{+0.65}_{-0.91}$, with a flat universe value of 
$\om = 0.32^{+0.07}_{-0.07}$.
The principal axes of this fit are almost identical to those of
the CMAGIC fit, so it is useful to express them in this frame.
Here they correspond to $\Omega_1 = -0.167^{+0.146}_{-0.133}$ and 
$\Omega_2 = 1.969^{+0.787}_{-1.146}$ (statistical errors only).
Note that the \mb\ fits agree somewhat better with a flat universe
than the \Bbvzs\ fits.

If \mb\ and \Bbvzs\ were equivalent (given current templates) we
would expect \al\ to be identical for the two methods.  When comparing
these numbers the marginalized, one-dimensional errors are appropriate 
instead of the outer extent of the 1 \sg\ error contours quoted 
previously. For \Bbvzs\ $\al = 0.516^{+0.193}_{-0.206}$,
and for \mb\ it is $\al = 0.995^{+0.253}_{-0.226}$, a difference
of 1.6 \sg .  They are marginally inconsistent, but not at a
significant level.

Directly comparing the \mb\ and \Bbvzs\ cosmological results
requires that the correlation between the two methods be measured,
and then propagated into the cosmological parameter space.
The details of this process are presented in 
Appendix~\ref{apndx:correlations}.  The result is that the 
correlation coefficients between the
two fits are 0.34 along the $\Omega_1$ axis and 0.15
along $\Omega_2$.

While many of the systematic errors should affect \mb\ and \Bbvzs\ 
equally, not all apply to both fits.  For example,
the number of points in the CMAGIC linear region is meaningless
in an \mb\ context.  Furthermore, individual SNe may have quite different
weights in the two fits, which partially removes the insensitivity
to systematics.  Both issues must be addressed before the results
can be compared.  The number of points in the linear region
and the detectability of CMAGIC bumps at high redshift have
already been discussed, and are summarized in
table~\ref{tbl:identifiedsystematics}.  In addition, we expect that
the effects of the $U-B$ color of the spectral templates
will not be the same for both methods, since \mb\ and \Bbvzs\ depend
on color information in a very different fashion.  Comparing the
results of \mb\ and \Bbvzs\ fits using the $U$-enhanced spectral templates
as discussed in \S\ref{subsec:systematicsfitting}, we find that
the residual difference due to this systematic is 
$\Delta \Omega_1 = 0.010$, $\Delta \Omega_2 = 0.151$.  The
effects of the differing weights can be addressed by performing a fit
to \mb\ where each SN is given the weight it has in the \Bbvzs\ fit,
and vice-versa.  It is not fair to include both values as systematics
errors, since they are essentially measuring the same effect.
Fortunately, they turn out to have almost identical effects.
The short axis is brought into better agreement by a
shift of $\Delta \Omega_1 = 0.054$ and the long axis by
$\Delta \Omega_2 = 0.31$.

Putting these contributions together, and using the correlations
given above, we find that the difference between the \mb\ and
\Bbvzs\ fits is
\begin{eqnarray*}
\Delta \Omega_1 & = & -0.182 \pm 0.097 \mbox{(stat)} 
 \pm 0.058 \mbox{(sys)} \\
\Delta \Omega_2 & = & 0.530 \pm 0.661 \mbox{(stat)}
 \pm 0.414 \mbox{(sys)}.
\end{eqnarray*}
The difference along the $\Omega_1$ axis amounts to 1.6 \sg ,
and along the $\Omega_2$ axis to 0.7 \sg . The major
disagreement is along the short axis, as is obvious from 
figure~\ref{fig:contcompare}, and a disagreement of this size or larger 
is expected to occur in 11\% of measurements.  Since $\Omega_1$ is 
essentially sensitive to acceleration, this amounts to the statement 
that the \Bbvzs\ results favor more acceleration at the 1.6 \sg\ level.  
The differences along both axes can be combined into one measure by 
projecting them along the difference vector, defined by 
$\Omega_3 \equiv -0.325 \Omega_1 + 0.946 \Omega_2$.  Then the difference 
between the two fits is $\Delta \Omega_3 = 0.560 \pm 0.657 \mbox{(stat)} \pm 
 0.410 \mbox{(syst)}$, a difference of 0.7 \sg .

\begin{figure}
\plotone{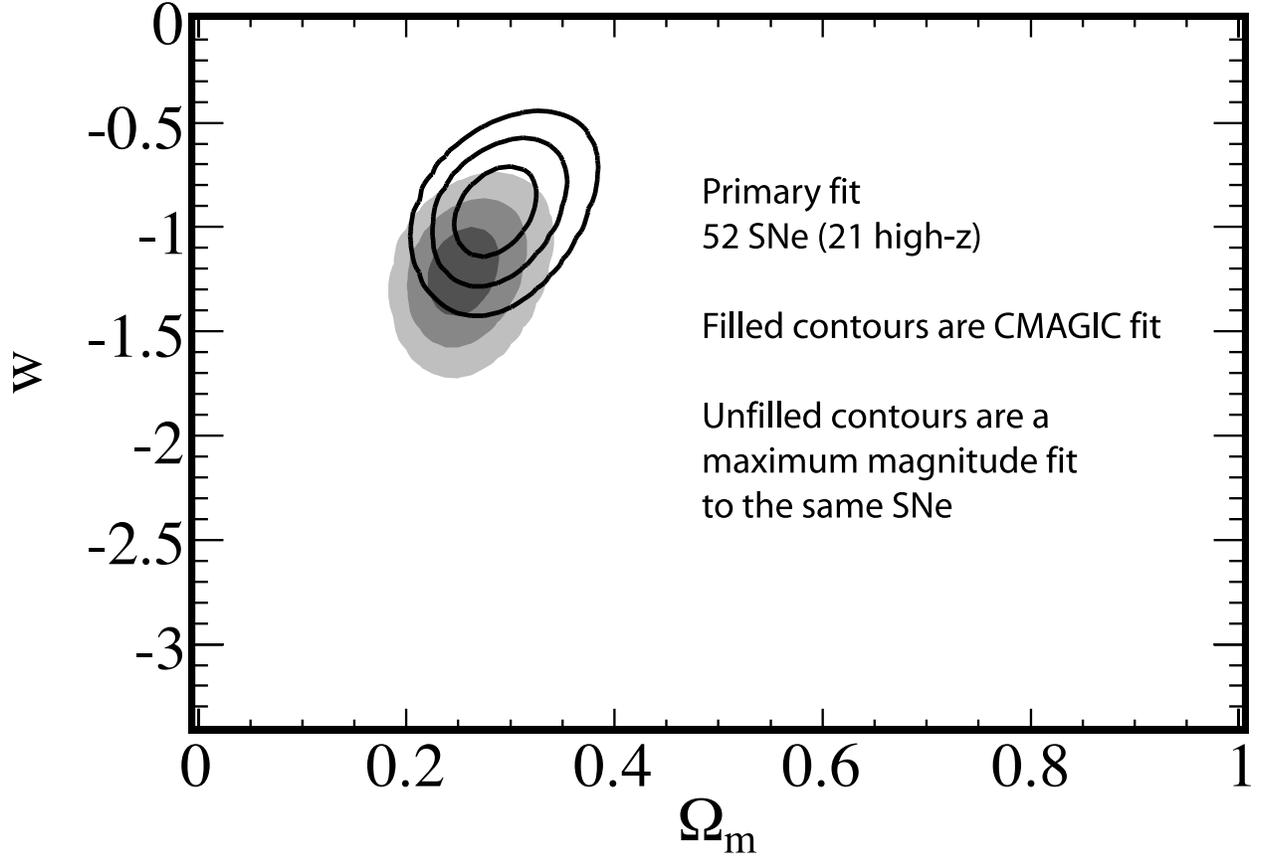}
\caption[Comparison of $w$ results for \mb\ and \Bbvzs]
{Direct comparison of the \mb\ and \Bbvzs\ cosmology fits
 to $w$, \om , assuming a flat universe and including the baryon acoustic peak
 measurement of \citet{Eisenstein:05}. The \mb\ results are shown as 
 unfilled contours and the \Bbvzs\ results as filled ones.  In both cases, the
 contours represent 68.3\%, 95.4\%, and 99.7\% of the total 
 probability. Note that the difference between the two results is
 certainly not independent of the difference between the \mb\ and \Bbvzs\
 results for \om\ and \ol . \label{fig:wcompare} }
\end{figure}

A similar comparison is possible with the \om , $w$ fits.
The result is shown in figure~\ref{fig:wcompare}.  The
same sort of detailed comparison is not carried out here
for several reasons.  First, the difference is certainly not
independent from the difference observed in \om , \ol\ space,
so little additional information would be gained from this
procedure.  Second, because the current constraints on
\om , $w$ from SN data alone are not well behaved
(not closing off until very negative values of $w$), it is not
useful to compare the two fits without the addition of additional
constraints, here the BAP measurement, which
is the same between the two fits.

\section{CONCLUSIONS}
\nobreak
CMAGIC provides some additional information that is not
captured by the standard light-curve template fitting techniques
used to estimate \mb .  This allows us to provide some additional
constraints on the cosmological parameters.  Furthermore, \Bbvzs\
should be affected differently by several potential evolutionary
effects.  

We have carried out the first blind analysis of the cosmological
parameters using SN data, developing a technique to prevent
experimenter bias by hiding the final result until the data cuts
and analysis procedures are finalized.  We
find that the results of a CMAGIC fit broadly confirm our picture
of an accelerating Universe.  In fact, they favor a higher amount of
acceleration than the \mb\ results by approximately 1.6 \sg\ (including
systematics and the correlations between the two measurements).
The \Bbvzs\ error contours differ from a flat Universe
by 1.7 \sg\ (including systematics), which would be interesting if it
were more statistically significant.

The constraints on the cosmological parameters from 
a CMAGIC fit to 31 nearby and 21 distant SNe~Ia are
$\om = 1.26^{+0.38}_{-0.51}$, $\ol = 2.20^{+0.41}_{-0.67}$ 
(statistical errors only).  However, this is a poor frame for
expressing the results.  It is significantly more useful to
instead quote the results as
\begin{displaymath}
\Omega_1 = 0.790 \om - 0.613 \ol =  -0.349^{+0.117}_{-0.131}
 \left( \mbox{stat} \right) ^{+0.060}_{-0.062} 
 \left( \mbox{syst} \right)
\end{displaymath}
\begin{displaymath}
 \Omega_2 = 0.613 \om + 0.790 \ol = 2.502^{+0.530}_{-0.838}
  \left( \mbox{stat} \right) ^{+0.476}_{-0.545} 
 \left( \mbox{syst} \right)
 \end{displaymath}
 with
 \begin{displaymath}
 \Omega_{m} = 0.19^{+0.06}_{-0.06}  \left( \mbox{stat} \right)
  ^{+0.03}_{-0.05} \left( \mbox{syst}\right)
\end{displaymath}
for a flat Universe, where the dark energy has been assumed to
have a constant equation of state with $w = -1$, as is the case for
a cosmological constant.  The systematic errors have been estimated by
considering a wide range of alternatives to the primary fit of this
paper.  The largest systematic error is the extinction cut,
indicating that while CMAGIC has some benefits with respect
to extinction by interstellar dust, we still have a great deal to
learn about this issue.  A direct comparison is also possible
with an \mb\ fit to the same SN, which requires that the correlations
between the two methods be estimated.  After including the systematics
and correlations, the difference between the two fits is almost exclusively
along the short axis, with the CMAGIC fits favoring more acceleration by
1.6\sg .  Fitting for a constant value of $w$ in a flat Universe,
the combination of the CMAGIC results  with the angular scale of
the BAP measured in \citet{Eisenstein:05} yields
$w = -1.21^{+0.15}_{-0.12} \left( \mbox{stat} \right)
 ^{+0.07}_{-0.12} \left( \mbox{supernova syst} \right)$,  
$\om = 0.25^{+0.02}_{-0.02} \left( \mbox{stat} \right)
 ^{+0.01}_{-0.01} \left( \mbox{supernova syst} \right)$, consistent 
with a cosmological constant at the $1.2 \sigma$ level.

The currently available high redshift SN
sample was not observed in an optimal fashion for CMAGIC.
Out of the approximately 100 published high-redshift SNe light curves,
only about 20 are useful for \Bbvzs .  As a result, the current data set
does not place strong constraints on dust or evolutionary effects.
This situation will change
in this decade; within the next 5 years it should be possible to
measure both \Bbvzs\ and \mb\ for 1000 high-redshift SNe, at which point the
comparison between \mb\ and \Bbvzs\ will be extremely interesting.

\acknowledgments

The authors would like to thank Brian Schmidt for providing non-$K$-corrected
light curves for SN 1997ce and SN 1997cj. This research has 
made use of the NASA/IPAC Extragalactic Database (NED) which is operated 
by the Jet Propulsion Laboratory, California Institute of Technology, 
under contract with the National Aeronautics and Space Administration.

\appendix
\section{CMAGIC BUMPS}
\label{apndx:bumps}
An example of an SNe~Ia with a bump feature is shown in the bottom
of figure~\ref{fig:cmag}.  Bumps seem to be associated with
SNe with different $B$ and $V$ stretches (where the templates have
been normalized such that the majority of SNe are well fitted with
the same $B$ and $V$ stretch), in particular when $s_V < s_B$.
In general, the probability of a bump increases with $B$ stretch.
It is possible to find examples of SNe~Ia with virtually the same stretch
but where one has a bump and the other does not.  This clearly
indicates that SNe~Ia do not constitute a one-parameter family,
at least in terms of stretch, \dmofb\ or the MLCS parameter $\Delta$. 
Bumps are far more common in other filter combinations.

However, these matters do not concern us here.  The
important thing for the purposes of this paper is the effect of the
bump on the cosmology fits.  As noted previously, the presence of
the bump has an effect on the starting and ending dates of the linear
feature.  With high-quality data it is trivial to detect the presence
of a bump.  Therefore, while this is not an issue with the low-redshift SNe,
it is a potential systematic in the cosmology fits due to the
lower quality of the high-redshift data making bumps difficult to detect
for some SNe.  Fortunately, this turns out to have a relatively small
effect for the present sample.

In order to quantify this effect, we attempted to determine 
the probability, as a function of stretch, that an SN has a bump
by examining the low redshift sample.  We find that all
SNe with $s > 1.1$ have a bump feature, and none with $s < 0.8$ do.
Between these extremes the probability of having a bump is an
increasing function of stretch, but remains probabilistic.
For $1.1 < s < 1.0$ approximately 50\% of SNe~Ia have bumps, and 
for $1.0 < s < 0.8$ only 1 out of 13 does.  Applying this result
to the high redshift sample, we see that there are six SNe in the first
group and 14 in the second.  One of the 14 (SN 1997ce) has a bump,
consistent with the predicted fraction.  As expected, individual
filter fits to SN 1997ce show that the $V$ stretch is less than the
$B$ stretch, with $s_B = 0.932 \pm 0.025$ and $s_V = 0.816 \pm 0.019$.
The systematic effect, if any,
will clearly arise from the first group, which consists of
SNe 1995ba, 1997F, 1998aw, 1999fj, 2001ix, and 2002ad.
The CMAGIC fits to SN 1998aw are not affected by the 
presence or absence of a bump, so it can be ignored for the purposes
of this discussion.

In order to quantify the probability that each of these SNe has an 
undetected bump, we analyzed a handful of very well observed low redshift 
SNe that have a bump feature (SNe 1995D, 1995bd, 1998bu, and 1999ee) 
and used their CMAGIC diagrams to quantify the excess $B$ magnitude over the 
value predicted by the CMAGIC linear fit as a function of rest frame epoch.
We then compared these values with the actual data 
points for the four high-redshift SNe in question, taking into account the 
observational errors and the dispersion of excess magnitudes in the bump.  
SN 1999fj, SN 2001ix, and SN 2002ad are inconsistent with a bump at greater
than the 2.5 \sg\ level.  No strong statement can be made for SN 1995ba 
or SN 1997F. Therefore, these are the only two that need concern us.

This gives four possibilities, which occur with approximately equal 
probability.  The case where neither has a bump is identical to our primary 
fit.  In order to estimate the systematic error associated with the other 
possibilities, we performed and compared all four fits, obtaining results
very similar to our primary fit.  We find that the effects of this systematic
on the current sample are
$\Delta \Omega_1 = 0.005$ and $\Delta \Omega_2 = -0.014$.
This indicates that undetected bumps do not contribute 
substantially to the systematic error.  The story is somewhat complicated,
but we have been fortunate in that it does not affect the current result.
Most future projects, which will obtain considerably
more complete color coverage, should not have to worry about this issue.

\section{CORRELATIONS BETWEEN THE \mb\ AND \Bbvzs\ FITS}
\label{apndx:correlations}
\nobreak
In order to determine the correlation between the cosmological
results of the \mb\ and \Bbvzs\ fits, it is first necessary to determine
the correlations between \mb\ and \Bbvzs\ values for each SN.
There are two components to this correlation: that induced
by the fitting procedures, and that intrinsic to the physics of
SNe~Ia and their environment (extinction, etc.).  The former can
be determined individually for each SN, and is seen to vary considerably
depending on the distribution of observations, while we are forced to
assume that the latter is constant across the SN sample.

\begin{figure}
\plotone{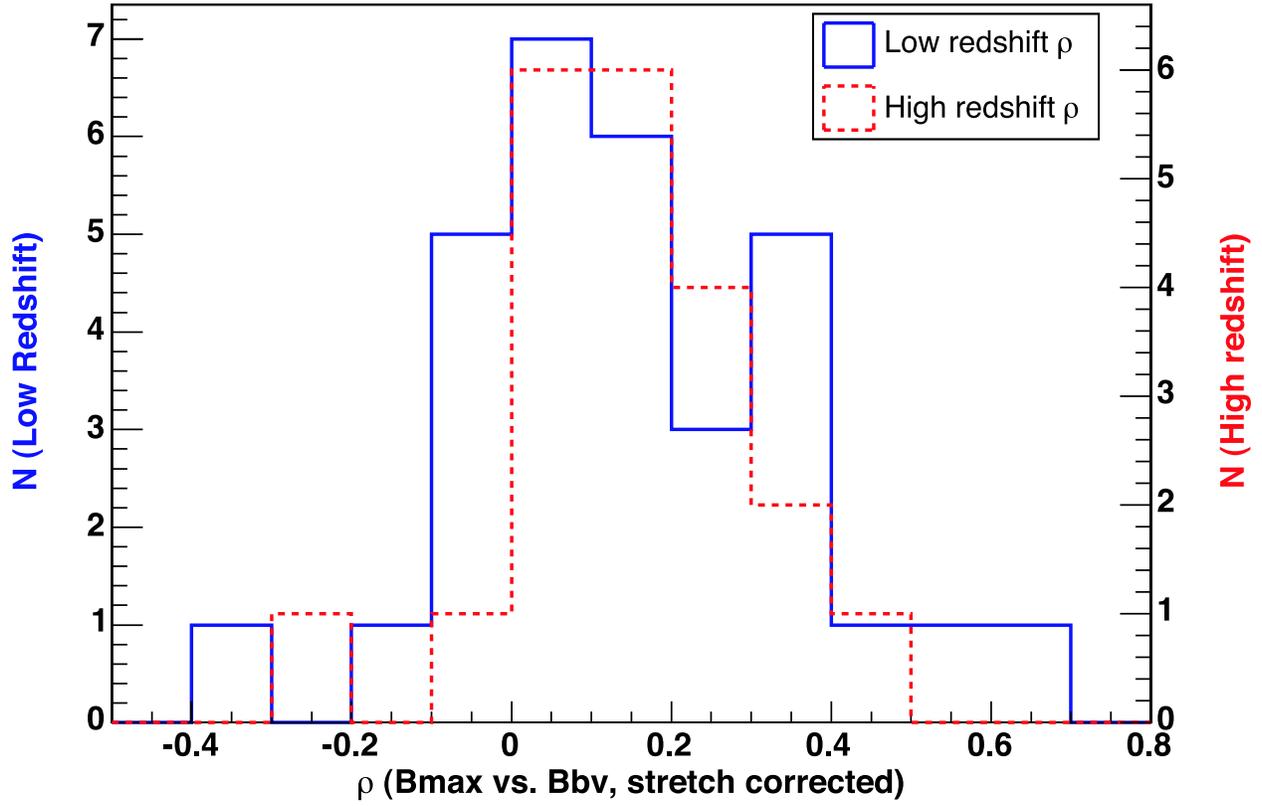}
\caption{ The histogram of correlation coefficients ($\rho$) for
 \mb , \Bbvzs\ after stretch correction.  This only includes the
 correlations induced by the fitting, and does not include any
 due to the intrinsic variability of SNe~Ia, or that which arises from
 peculiar velocity errors. \label{fig:magcorr} }
\end{figure}

Since current light-curve templates do not adequately reproduce
the CMAGIC relations, the fit correlation must be determined by a
Monte Carlo process.  For every SN, 1000 realizations are generated
using the actual photometric errors and observed epochs.  For each
realization \mb\ and \Bbvzs\ are fitted independently, and the correlations
are estimated from the resulting distributions.  After stretch
correction, the correlation between \mb\ and \Bbvzs\ is small
and positive, with mean correlation coefficients of 
$\left< \rho \right> = 0.150$ at 
low redshift and $\left< \rho \right> = 0.144$ for distant SNe.
The distributions are shown in figure~\ref{fig:magcorr}.  Furthermore,
the correlation between stretch and \Bbvzs\ is quite weak,
justifying the assumption that they are uncorrelated in the CMAGIC
fitting procedure ($\left< \rho \right> = 0.097$). 

\begin{figure}
\plotone{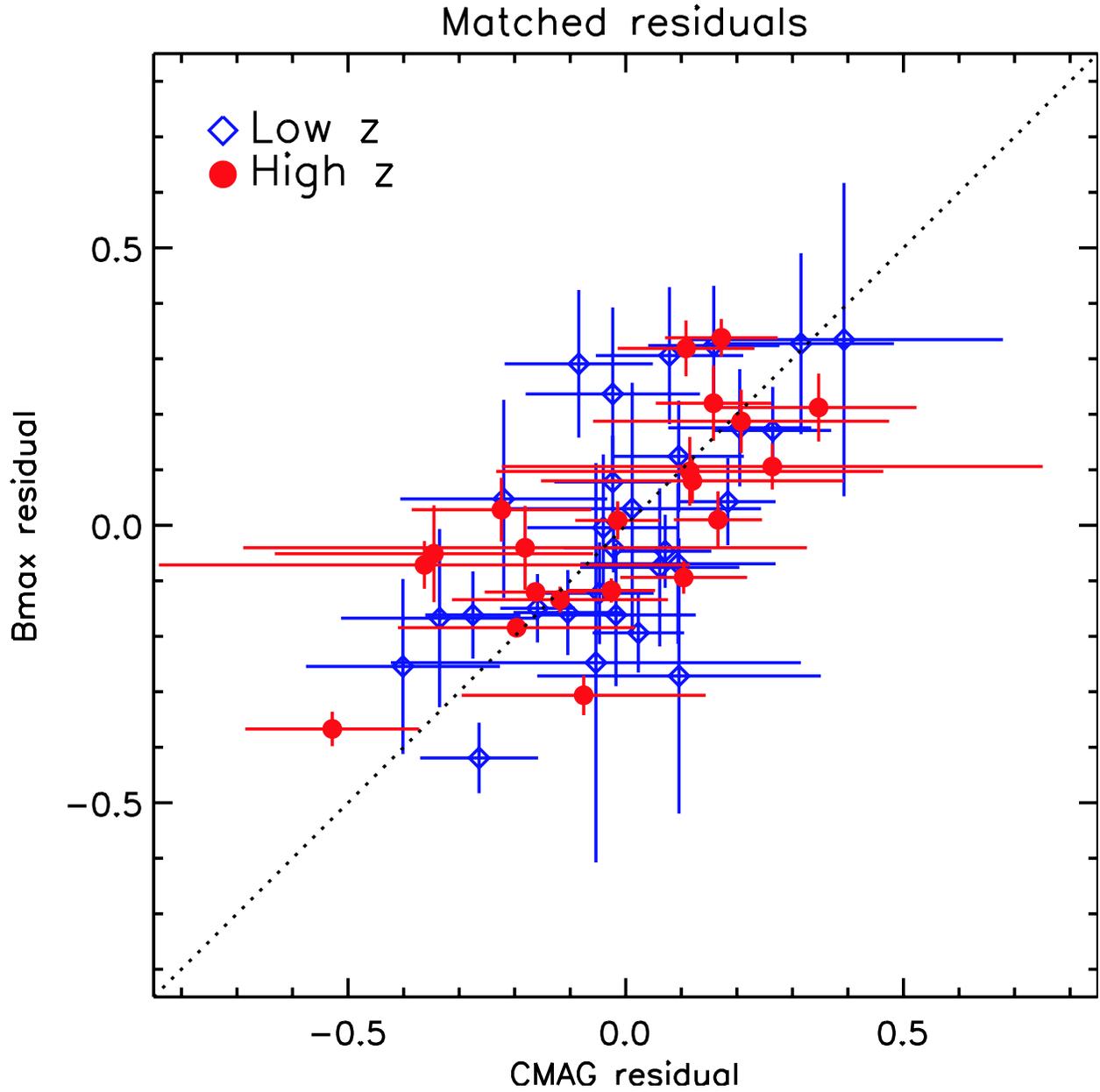}
\caption{Residuals from Hubble line for the \mb\ fit
and the \Bbvzs\ fit.  Overlain is a line with slope 1.  The high
redshift data is shown as filled circles and the low redshift data as
open diamonds.  The correlation of the low redshift sample is mostly
explained by the effects of peculiar velocities. \label{fig:residscatter}}
\end{figure}

In order to estimate the residual resulting from the intrinsic
heterogeneity of SNe~Ia, the best tool is to consider the residual
versus residual plot, shown in figure~\ref{fig:residscatter}.
Note that these residuals are with respect to different fits
with different values of the cosmological parameters.
It is clear that they are correlated, although this is much less true
of the high redshift sample:
$\mbox{cov}\! \left[ \mb , \Bbvzs \right]_{\mbox{lowz} } = 0.020$ and
$\mbox{cov}\! \left[ \mb , \Bbvzs \right]_{\mbox{highz} } = 0.0076$,
where cov denotes the covariance between the two quantities.
These values correspond roughly to $\rho = 0.55$ and $\rho = 0.34$,
respectively.  It is not surprising
that the low-redshift sample shows considerably more
correlation because of the dominant role of peculiar
velocity errors, which affect \mb\ and \Bbvzs\ identically.

To estimate the intrinsic correlation it is necessary to
subtract the effects of both the peculiar velocity and the
correlations induced by the light-curve and CMAGIC fitting
procedures.  If $r_{\mb}$ and $r_{\Bbvzs}$ denote the residuals from the
fit, then, using the low-redshift approximation for ${\mathcal D}_L$
(which is appropriate because peculiar velocities have a negligible
effect at high redshift),
and noting that the stretch and redshift are anti-correlated, 
\begin{eqnarray}
 \mbox{cov}\! \left[ r_{\mb} , r_{\Bbvzs} \right] & = &
  \mbox{cov}\! \left[ \mb , \Bbvzs \right] + 
  \alpha_{\mb} \mbox{cov}\! \left[ s , \Bbvzs \right] + \\
  \nonumber & \nonumber & 
  \alpha_{\Bbvzs} \mbox{cov}\! \left[ s, \mb \right] + 
  \sigma^2_s \alpha_{\mb} \alpha_{\Bbvzs} +
  \left( \frac{5}{\log 10} \frac{\sigma_z}{z} \right)^2 + \\
  \nonumber & \nonumber &
  \frac{5}{\log 10} \left( \alpha_{\mb} + \alpha_{\Bbvzs} \right) 
  \frac{ \sigma_z } { z } \sigma_s +
  \mbox{cov}\! \left[ \scriptm_{\mb}, \scriptm_{\Bbvzs} \right] .
\end{eqnarray}
Here the correlations between stretch, \mb , and \Bbvzs\ are those
arising from the fitting procedure only. The desired quantity is 
$\mbox{cov}\! \left[ \scriptm_{\mb}, \scriptm_{\Bbvzs} \right]$,
the correlation between the absolute magnitudes modulo the
Hubble constant.  Note
that the stretch-corrected covariance shown in figure~\ref{fig:magcorr}
is not appropriate here because the contributions of stretch are
handled separately.  More than half of the measured 
covariance in the low redshift sample (0.013) comes from peculiar
velocity errors, which have essentially no effect on the high
redshift sample.  We find that 
$\mbox{cov}\! \left[ \scriptm_{\mb}, 
 \scriptm_{\Bbvzs} \right]_{\mbox{lowz}} =  0.0072 $ and
$\mbox{cov}\! \left[ \scriptm_{\mb}, 
 \scriptm_{\Bbvzs} \right]_{\mbox{highz}} = 0.0044$,
which correspond to $\rho = 0.37 \pm 0.14 $ and $\rho = 0.22 \pm 0.21$,
respectively.  These are consistent, and therefore the
overall correlation coefficient for the intrinsic scatter
is taken to be $\rho = 0.32 \pm 0.12$.

Next it is necessary to propagate this covariance into
the cosmological parameter space.  This is far
from straightforward.  While it might be tempting to
simply assume that the intrinsic correlation is the
dominant one, and that this can therefore be taken
as the correlation between the $\Omega_1$ values
of the two fits, there is no way to justify this
assumption.  The correlation at low redshift
is dominated by the peculiar velocity errors, and it
is unclear how important this is in the context of
the cosmological parameters.  Furthermore, different
SNe have different weights, both because of their observational errors 
and because SNe at different redshifts have different influences
in the \om , \ol\ parameter space.

In order to determine the effects of these correlations
on $\Omega_1$, $\Omega_2$, a Monte-Carlo simulation
was carried out on the SN samples.  The covariances between 
stretch, \mb\ and \Bbvzs\ from the fitting procedures were 
calculated for each supernova as described above, to which 
were added the measured intrinsic correlation coefficient of 
0.32.  This simulation also incorporated the effects of redshift errors 
including the assumed peculiar velocity of 300 km s$^{-1}$.  

Generating 2500 realizations required approximately 4 days
on a fast workstation.  The corresponding correlation
coefficients for the $\Omega_1$ and $\Omega_2$ axes are
$\rho_{11} = 0.34 \pm 0.02$ and $\rho_{22} = 0.15 \pm 0.02$.
The correlation is not evenly distributed 
between the two axes, acting primarily along the short axes of the 
error ellipses. Since these correlations are positive, they act to 
increase the significance of the difference between the two fits.
The same data set can be used to verify that 
$\Omega_1$ and $\Omega_2$ are uncorrelated, yielding 
$\rho_{ 12\mb } = -0.07$ and $\rho_{ 12\Bbvzs } = 0.07$.

\section{SUPERNOVAE REMOVED BY EACH CUT}
\label{apndx:cutremoved}
\nobreak
This section presents a list of the SNe removed by each cut at
the values specified in the primary fit.  Note that these cuts
are not applied in any order, and therefore some SNe fail multiple cuts.
Furthermore, some of the cuts are correlated.  For example, an SN that
does not have any data within 7 days of $B$ maximum is unlikely to have
a well-determined date of maximum.

The following SNe do not have any points in their CMAGIC linear region:
SNe 1995ar, 1995aw, 1995ay, 1995az, 1996cf, 2001iw, and 2002P.
These were at redshifts too low to be used in the cosmology fit (although
some were used to determine the intrinsic \betabv\ distribution):
SN1990N, SN1994ae, SN1995D, SN1995al, SN1996X, SN1996Z, SN1997bp,
SN1997bq, SN1997br, SN1998bu, SN1998dh, SN1999ac, SN1999by, SN1999cl,
SN1999gh, SN2000E, SN2001el, SN2002bo.  The following SN did not
have data with 7 rest frame days of $B$ maximum: 
SN1990T, SN1990Y, SN1991S, SN1991U,
SN1991ag, SN1992bg, SN1992bk, SN1993ae, SN1993ah, SN1994Q, SN1997bq,
SN1998ec, SN1999gh, SN2000bh, SN2000ce, SN2001jn, SN2001jp, and SN2002P.
These SNe did not have a well determined date of maximum:
SN1992ae, SN1992au, SN1992bk, SN1993B, SN1993ah, SN1994G, SN1995ak,
SN1995aq, SN1995ar, SN1995ax, SN1995ay, SN1996I, SN1996U, SN1996Z,
SN1996cm, SN1997K, SN1997S, SN1999fn, SN2000bh, SN2001hx, SN2001hy,
SN2001jb, SN2001jf, SN2001jn, and SN2002P.
The following SNe had stretch values below the
minimum cutoff, and were removed for the reasons discussed in
\S \ref{subsec:cuts}: SN1992au, SN1998bp, SN1998de, SN1999by.  
SN1996U, SN1997K, SN1997am, SN1999ff,
SN2001hx, SN2001hy, and SN2001jb have errors on \Bbvzs\ that exceeded
0.5 mag.  SN1990Y, SN1992J, SN1993H, SN1995E, SN1995bd, SN1996C, 
SN1996Z, SN1996bo,
SN1997br, SN1998aw, SN1998bu, SN1999cl, SN1999ee, SN1999fw,
SN1999gd, SN2000ce, SN2001jn, SN2002ad, and SN2002bo have measured color 
excesses larger than the 0.25 mag cut value.  
As discussed in \S \ref{subsec:cuts}, there were
3 additional SN removed by hand from the sample for various reasons:
SN1997O, SN1997br, and SN1997bp.

\clearpage

\begin{deluxetable}{llllr}
\tabletypesize{ \small }
\tablecaption{Low redshift SNe used in primary 
 fit\label{tbl:primarysamplowz} }
\tablehead{ 
 \colhead{IAU Name} & \colhead{  $z_{hel}$ \tablenotemark{a} } &
 \colhead{ stretch } & \colhead{ \Bbvzs \tablenotemark{b}} &
 \colhead{Reference}
}
\startdata
SN1990O  & $ 0.031 $ & $ 1.087(032) $ & $ 17.530(067) $ & 1 \\
SN1990af & $ 0.050 $ & $ 0.750(010) $ & $ 18.894(078) $ & 1 \\
SN1992ag & $ 0.026 $ & $ 0.959(022) $ & $ 17.222(056) $ & 1 \\
SN1992al & $ 0.014 $ & $ 0.929(013) $ & $ 15.838(083) $ & 1 \\
SN1992bc & $ 0.020 $ & $ 1.079(007) $ & $ 16.738(062) $ & 1 \\
SN1992bh & $ 0.045 $ & $ 1.057(024) $ & $ 18.697(050) $ & 1 \\
SN1992bl & $ 0.043 $ & $ 0.845(021) $ & $ 18.556(065) $ & 1 \\
SN1992bo & $ 0.018 $ & $ 0.744(007) $ & $ 16.918(049) $ & 1 \\
SN1992bp & $ 0.079 $ & $ 0.897(021) $ & $ 19.634(086) $ & 1 \\
SN1992bs & $ 0.063 $ & $ 1.025(017) $ & $ 19.568(071) $ & 1 \\
SN1993O  & $ 0.052 $ & $ 0.927(020) $ & $ 18.912(034) $ & 1 \\
SN1993ag & $ 0.049 $ & $ 0.940(027) $ & $ 18.839(064) $ & 1 \\
SN1994M  & $ 0.023 $ & $ 0.883(025) $ & $ 17.422(084) $ & 2 \\
SN1994S  & $ 0.015 $ & $ 1.052(024) $ & $ 16.181(102) $ & 2 \\
SN1996bl & $ 0.036 $ & $ 1.014(014) $ & $ 17.879(029) $ & 2 \\
SN1996bv & $ 0.017 $ & $ 1.039(020) $ & $ 16.225(030) $ & 2 \\
SN1997E  & $ 0.013 $ & $ 0.821(006) $ & $ 16.232(038) $ & 3 \\
SN1998V  & $ 0.018 $ & $ 0.962(040) $ & $ 16.389(068) $ & 3 \\
SN1998ab & $ 0.027 $ & $ 0.958(006) $ & $ 17.212(036) $ & 3 \\
SN1998es & $ 0.011 $ & $ 1.075(014) $ & $ 15.074(048) $ & 3 \\
SN1999aa & $ 0.014 $ & $ 1.098(004) $ & $ 16.135(017) $ & 4 \\
SN1999aw & $ 0.038 $ & $ 1.358(008) $ & $ 18.242(035) $ & 5 \\
SN1999dk & $ 0.015 $ & $ 1.089(010) $ & $ 15.862(020) $ & 6 \\
SN1999dq & $ 0.014 $ & $ 1.060(004) $ & $ 15.498(076) $ & 3 \\
SN1999ek & $ 0.018 $ & $ 0.895(007) $ & $ 16.573(049) $ & 7 \\
SN1999gp & $ 0.027 $ & $ 1.141(004) $ & $ 17.222(064) $ & 6 \\
SN2000ca & $ 0.024 $ & $ 1.007(016) $ & $ 17.137(067) $ & 7 \\
SN2000dk & $ 0.017 $ & $ 0.720(004) $ & $ 16.394(037) $ & 3 \\
SN2000fa & $ 0.021 $ & $ 0.972(007) $ & $ 17.025(062) $ & 3 \\
SN2001V  & $ 0.015 $ & $ 1.119(017) $ & $ 15.769(110) $ & 8 \\
SN2001ba & $ 0.029 $ & $ 1.049(014) $ & $ 17.669(042) $ & 7 \\
\enddata
\tablenotetext{a}{ Heliocentric redshift. }
\tablenotetext{b}{ Does not include \sgi }
\tablecomments{SNe in primary cosmology fit, not including SNe not in the
Hubble flow used to measure the slope distribution.}
\tablerefs{ (1) \citet{Hamuy:96}, (2) \citet{Riess:99a}, (3) \citet{Jha:05},
 (4) \citet{Krisciunas:00}, (5) \citet{Strolger:02}, (6) \citet{Krisciunas:01},
  (7) \citet{Krisciunas:04}, (8) \citet{Vinko:03}}
\end{deluxetable}

\begin{deluxetable}{llllr}
\tabletypesize{\small}
\tablecaption{High redshift SNe used in primary 
 fit \label{tbl:primarysamphighz} }
\tablehead{ 
 \colhead{IAU Name} & \colhead{  $z_{hel}$ \tablenotemark{a} } &
 \colhead{ stretch } & \colhead{ \Bbvzs \tablenotemark{b}} &
 \colhead{Reference}
}
\startdata
SN1995K  & $ 0.479 $ & $ 0.956(046) $ & $ 24.276(220) $ & 1 \\
SN1995ba & $ 0.388 $ & $ 0.999(052) $ & $ 24.025(267) $ & 2 \\
SN1996E  & $ 0.430 $ & $ 0.940(005) $ & $ 23.572(156) $ & 3 \\
SN1996K  & $ 0.380 $ & $ 0.888(013) $ & $ 24.169(166) $ & 3 \\
SN1997F  & $ 0.580 $ & $ 1.034(070) $ & $ 24.861(349) $ & 2 \\
SN1997H  & $ 0.526 $ & $ 0.883(051) $ & $ 24.242(478) $ & 2 \\
SN1997P  & $ 0.472 $ & $ 0.898(039) $ & $ 24.610(487) $ & 2 \\
SN1997ai & $ 0.450 $ & $ 0.918(112) $ & $ 23.876(283) $ & 2 \\
SN1997af & $ 0.579 $ & $ 0.846(050) $ & $ 24.655(508) $ & 2 \\
SN1997ce & $ 0.440 $ & $ 0.932(025) $ & $ 24.327(062) $ & 4 \\
SN1997cj & $ 0.500 $ & $ 0.925(021) $ & $ 24.453(077) $ & 4 \\
SN1997eq & $ 0.540 $ & $ 0.947(026) $ & $ 24.514(194) $ & 5 \\
SN1998as & $ 0.355 $ & $ 0.961(023) $ & $ 23.786(100) $ & 5 \\
SN1998ax & $ 0.497 $ & $ 1.156(032) $ & $ 24.447(115) $ & 5 \\
SN1998ba & $ 0.430 $ & $ 0.975(022) $ & $ 24.241(091) $ & 5 \\
SN1999fj & $ 0.816 $ & $ 1.037(040) $ & $ 25.517(273) $ & 6 \\
SN2000fr & $ 0.543 $ & $ 1.100(020) $ & $ 24.542(079) $ & 5 \\
SN2001iv & $ 0.397 $ & $ 0.977(004) $ & $ 23.720(091) $ & 7 \\
SN2001ix & $ 0.711 $ & $ 1.025(052) $ & $ 24.937(159) $ & 7 \\
SN2002ab & $ 0.423 $ & $ 0.924(015) $ & $ 23.872(214) $ & 7 \\
SN2002kd & $ 0.735 $ & $ 0.907(013) $ & $ 25.385(114) $ & 8 \\
\enddata
\tablenotetext{a}{ Heliocentric redshift. }
\tablenotetext{b}{ Does not include \sgi }
\tablecomments{SNe in primary cosmology fit, not including SNe not in the
Hubble flow used to measure the slope distribution.}
\tablerefs{ (1) \citet{Schmidt:98}, (2) \citet{Perlmutter:99}, (3) 
 \citet{Riess:98}, (4) B.\ Schmidt, private communication, (5) 
 \citet{Knop:03}, (6) \citet{Tonry:03}, (7) \citet{Barris:04}, (8) 
 \citet{Riess:04} }
\end{deluxetable}

\begin{deluxetable}{lrcrcrrr}
\tablecaption{Best fit CMAGIC slopes for well observed high redshift SNe
 \label{tbl:highzslopes} }
\tablehead{
 \colhead{IAU Name} & \colhead{ $z$ } & \colhead{$N_{lin}$ \tablenotemark{a}} &
 \colhead{\chisq} & \colhead{DOF \tablenotemark{b}} & \colhead{Prob 
   \tablenotemark{c}} & \colhead{ \betabv } & \colhead{ $\sigma_{\betabv}$ }
}
\startdata
SN1997ce & 0.44  & 4 & 1.41 & 2 & 0.492 & 1.803 & 0.180 \\
SN1997cj & 0.5 & 6 & 1.44 & 4 & 0.839 & 2.159 & 0.264 \\
SN1998aw & 0.44  & 3 & 0.14 & 1 & 0.705 & 2.027 & 0.390 \\
SN1998ax & 0.497 & 3 & 0.43 & 1 & 0.513 &1.616 & 0.291 \\
SN1998ba & 0.43  & 3 & 0.004 & 1 & 0.952 & 2.222 & 0.359 \\
\enddata
\tablenotetext{a}{ Number of points in CMAGIC linear region.}
\tablenotetext{b}{Degrees of freedom ($N_{lin}-2$) for this fit.}
\tablenotetext{c}{Probability that the \chisq\ should be worse than
 the observed value.  A high number indicates the fit is `too good.'}
 \tablecomments{ Results of CMAGIC fits to well observed high redshift SNe.
  Unlike the fits used in the cosmological analysis, here the stretch \betabv\
  is fit for each SN. }
 \end{deluxetable}

\begin{deluxetable}{llll}
\tabletypesize{\small}
\tablecaption{Cuts and parameters for primary fit}
\tablehead{
 \colhead{Description} & \colhead{Value} & \colhead{Formal name}
  \label{tbl:primarycuts} }
 \startdata  
 Minimum redshift cutoff for cosmology fit & 0.01 & zmin \\
 Maximum redshift cutoff for cosmology fit & NA & zmax \\
 High redshift cutoff for slope distribution fit & 0.1 & zslopemax \\
 Minimum number of points in CMAGIC linear region & 1 & npointsmin \\
 Maximum allowable magnitude error & 0.5 mag & magerror \\
 Maximum allowable \BmV\ excess at $B_{max}$ & 0.25 mag & maxcolor\\
 Maximum allowable error in date of $B$ maximum & 1.0 days & datemaxerror \\
 Minimum stretch allowed & 0.7 & stretchmin \\
 Maximum gap between maximum and nearest point in $B$ or $V$ & 7 days & 
  daygap\\
 \enddata
 \tablecomments{ Cuts and their values for the 
 primary fit.   A cut that is not used in the primary fit is given a value 
 of NA.}
\end{deluxetable}

\begin{deluxetable}{lrrr}
\tabletypesize{\small}
\tablecaption{Identified Systematic Errors}
\tablehead{ \colhead{Source} & \colhead{$\Delta \Omega_1$ \tablenotemark{a}} &
 \colhead{$\Delta \Omega_2$ \tablenotemark{b}} & 
 \colhead{$\Omega_m$ (flat) \tablenotemark{c}}
 \label{tbl:identifiedsystematics} 
}
\startdata
{\bf Variation of fitting procedures } \\
\hline
\ No stretch correction\tablenotemark{d} 
 & $ -0.009 \left( 0.07 \sigma \right)$ &
 $ 0.060 \left( 0.11 \sigma \right) $ & $ -0.006 \left( 0.10 \sigma \right)$\\
\ P99 lightcurve fit & $ 0.015 \left( 0.13 \sigma \right) $ &
 $ 0.144 \left( 0.27 \sigma \right) $ & $ -0.007 \left( 0.12 \sigma \right)$\\
\ U-enhanced \kcorr\ & $-0.052 \left( 0.39 \sigma \right)$ &
 $ 0.100 \left( 0.19 \sigma \right) $ & $ -0.040 \left( 0.70 \sigma \right)$\\
{\bf Variation of cuts } \\
\hline
\ daygap $< 5$ & $ -0.024 \left( 0.18 \sigma \right) $ &
 $ 0.139 \left( 0.26 \sigma \right) $ & $-0.006 \left( 0.11 \sigma \right) $\\
\ daygap $< 10$ & $ 0.024 \left( 0.20 \sigma \right) $ &
 $ 0.015 \left( 0.03 \sigma \right) $ & $0.014 \left( 0.24 \sigma \right) $\\
\ $z > 0.015$ & $ -0.003 \left( 0.02 \sigma \right) $ &
 $ 0.023 \left( 0.04 \sigma \right) $ & $ 0.00 \left( 0.00 \sigma \right) $\\
\ magerror $< 0.25$ & $-0.011 \left( 0.08 \sigma \right)$ & 
 $ 0.015 \left( 0.03 \sigma \right) $ & $ -0.006 \left( 0.11 \sigma \right)$\\
\ magerror $< 1.0$ & $-0.011 \left( 0.09 \sigma \right)$ &
 $ 0.017 \left( 0.03 \sigma \right) $ & $ -0.006 \left( 0.11 \sigma \right)$\\
\ $E\left( B-V \right) < 0.1$ & $ 0.018 \left( 0.15 \sigma \right)$ &
 $ -0.467 \left( 0.56 \sigma \right)$ & $0.004 \left( 0.07 \sigma \right)$\\
\ $E\left( B-V \right) < 0.5$ & $ -0.048 \left( 0.36 \sigma \right) $ &
 $ -0.00 \left( 0.00 \sigma \right)$ & $0.017 \left( 0.29 \sigma \right)$\\
\ datemaxerror $<0.5$ & $ 0.005 \left( 0.04 \sigma \right) $ &
 $ 0.027 \left( 0.05 \sigma \right)$ & $0.00 \left( 0.00 \sigma \right)$\\
\ datemaxerror $<2$ & $0.018 \left(0.15 \sigma\right)$ &
 $ 0.115 \left(0.22 \sigma \right)$ & $0.017 \left( 0.29 \sigma \right)$ \\
\ npointsmin $>2$ & $0.018 \left(0.15 \sigma \right)$ &
 $0.229 \left( 0.43 \sigma \right)$ & $-0.006 \left( 0.11 \sigma \right)$\\
{\bf Other systematics } \\
\hline
\ Hamuy, Riess, Jha only & $-0.015 \left( 0.11 \sigma \right)$ &
 $ -0.037 \left( 0.04 \sigma \right)$ & $ -0.006 \left( 0.11 \sigma \right)$\\
\ Jack-Knife: SN2001ix & $ -0.014 \left( 0.11 \sigma \right) $ &
 $ -0.281 \left(0.33 \sigma \right) $ & $ -0.013 \left( 0.23 \sigma \right)$\\
\ Jack-Knife: SN2002kd & $ 0.015 \left( 0.12 \sigma \right) $ & 
 $ 0.310 \left( 0.58 \sigma \right) $ & $ -0.016 \left( 0.28 \sigma\right) $ \\
 \ $\sgi=0.08$ & $-0.015 \left(0.11 \sigma\right)$ &
 $0.022 \left(0.04 \sigma \right)$ & $-0.013 \left( 0.23 \sigma \right)$\\
\ $\sgi=0.15$ & $0.010 \left(0.09 \sigma\right)$ & 
 $-0.015 \left(0.02 \sigma\right)$ & $0.006 \left( 0.11 \sigma \right)$\\
\ Malmquist bias & $-0.0012 \left(0.01 \sigma\right)$ &
 $0.045 \left( 0.08 \sigma \right)$ & $-0.003 \left( 0.05 \sigma \right)$\\
\ Bumps & $ 0.005 \left( 0.04 \sigma \right) $ & 
 $-0.014 \left( 0.017 \right)$ & $0.003 \left( 0.05 \sigma \right)$\\
\enddata
\tablenotetext{a}{Shift in $\Omega_1 \equiv 0.790 \om - 0.613 \ol$ (the
 short axis).}
\tablenotetext{b}{Shift in $\Omega_2 \equiv 0.613 \om + 0.790 \ol$ 
 (the long axis).}
\tablenotetext{c}{Shift in the value of \om\ assuming a flat universe
 ($\om + \ol = 1$)}
\tablenotetext{d}{Not used in final determination of systematic errors.}
\tablecomments{Identified systematic errors, as detailed in 
 \S \ref{sec:systematics}.  The names of the cuts referenced above are
 as defined in table~\ref{tbl:primarycuts}. }
\end{deluxetable}

\clearpage

\epsscale{0.7}
\epsscale{1}

\epsscale{0.45}
\epsscale{1.0}

\epsscale{0.6}
\epsscale{1}

\epsscale{0.8}
\epsscale{1}

\epsscale{0.6}
\epsscale{1}

\end{document}